\documentclass[10pt,a4paper]{article}
\pdfoutput=1
\usepackage[english]{babel}
\usepackage{longtable}
\usepackage{amsmath,amsfonts,amssymb,bm}\interdisplaylinepenalty=2500

\newcommand{\unit}[1]{\ensuremath{\mathrm{#1}}}

\newcommand{\req}[1]{(\ref{#1})}
\usepackage{graphicx,color,rotating,epstopdf}
\graphicspath{{figs/}}					
\def\PrintGraphicFileName{1}			
\newcommand{\namedgraphics}[3]{
	\parbox{#3}{%
	\ifnum\PrintGraphicFileName>0\rotatebox{90}{\smash{\ttfamily\scriptsize\raisebox{0.8em}{#2}}}\fi%
	\hspace*{\fill}\includegraphics[scale=#1]{#2}\hspace*{\fill}}}

\newcommand{\ToDay}{\today}
\newcommand{\TODAY}{\today}
\title{Primary Calibration of AM and PM Noise Measurements}
\author{Enrico Rubiola\\
\small web page \texttt{http://rubiola.org}
\\[4em]\includegraphics[width=0.35\textwidth]{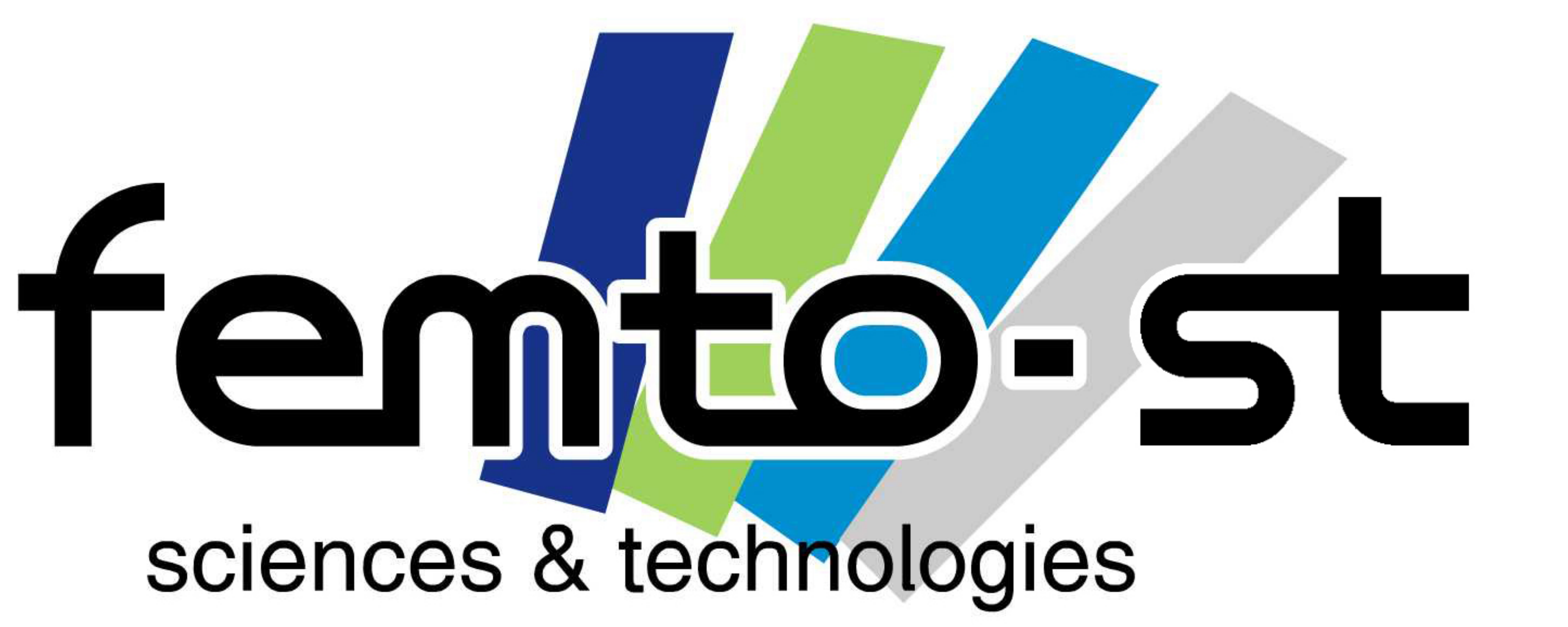}\\[0.5em]
\small FEMTO-ST Institute\\[-0.5ex]
\small CNRS and Universit\'e de Franche Comt\'e, 
\small Besan\c{c}on, France\\[1.5em]}
\date{\small\TODAY}
\begin{document}
\maketitle

\begin{abstract}
This report describes a method for the primary calibration of phase noise and amplitude noise measurement systems.

In the field of metrology, the term \emph{primary} refers to a standard whose quantity value and measurement uncertainty are established without relation to another measurement standard for a quantity of the same kind; or to a procedure used to realize the definition of a measurement unit and obtain the quantity value and measurement uncertainty of a primary measurement standard. 
Accordingly, a phase and a normalized amplitude have to rely on ratio measurements.

This is a working version, which I expect to update in the following six months adding some crucial experimental results. 

\end{abstract}
\clearpage
\tableofcontents

\clearpage
\section{Introduction}\label{sec:cal-introduction}
As physics and technology progress, phase noise becomes an increasingly relevant issue in a number of fields.  The demand for lower phase noise is seen in a number of examples, like high speed electronics, and telecommunications
Extreme low phase noise specifications are found in military and space oscillators.
Some types of military radars can achieve the desired resolution only if the phase noise of the X band reference signal does not exceed $10^{-14}$ \unit{rad^2/Hz} ($-140$ dB) at 10 kHz off the carrier.  Space research requires a frequency stability in the lower $10^{-15}$ at $\tau=1\ldots10^3$ s, which means a phase noise exceed $10^{-13}$ \unit{rad^2/Hz} ($-130$ dB) at 1 Hz off the carrier if the signal is taken at the standard 100 MHz output.
Of course, we expect that the need for measurement accuracy will inevitably follow the specifications for low phase noise, and consequently for high measurement sensitivity.
Besides, primary metrology is around the corner.  In fact, in 2010 the definition of the second is expected to switch\footnote{This was said at the panel discussion on the future of SI at the CPEM conference, Torino, Italy, July 9--14 2006.  Yet, it seems that this decision is still not in the official documents (thanks to Elio Bava, head of the INRIM, Torino).} from the $F\!=\!3 \leftrightarrow F\!=\!4$, $\Delta m_f=0$ hyperfine transition of the \unit{^{133}Cs} atom (9.192\,631\,770 GHz) to an optical transition, to be determined.  In this business, we expect that the phase noise of some synthesizers and transfer oscillators is a part of the metrological chain, for it has to be understood with high accuracy. 

This article privileges phase noise vs.\ amplitude noise because of the needs and of the culture of our laboratory.  Nonetheless, the method proposed here relies upon amplitude noise measurements, and provides a solution in which AM and PM calibration have the same accuracy.

In the field of metrology, the term \emph{primary} \cite{vim04iso,howarth00euromet,quinn97metro} refers to a standard whose quantity value and measurement uncertainty are established without relation to another measurement standard for a quantity of the same kind; or to a procedure used to realize the definition of a measurement unit and obtain the quantity value and measurement uncertainty of a primary measurement standard.  By contrast, \emph{secondary} means that the standard, or to the measurement procedure, relies upon calibration against, or comparison with, a primary measurement standard for a quantity of the same kind.
It is worth mentioning that an empirical unit can have higher reproducibility than the primary counterpart, as it happens with the Josephson and with the quantum-Hall electrical standards, yet inferior absolute accuracy as related to the SI\@.

\subsection{Basic definitions} 
The SI unit of angle, the \emph{radian} (rad) is now considered a \emph{derived unit} because an angle can always be defined in terms of the ratio of two homogeneous quantities.  
Formerly, it was considered an auxiliary unit.
In the domain of electric circuits, it is usual to represent the sinusoidal signal $v(t)=\sqrt{2}V_\text{rms}\cos[\omega_0t+\varphi]$ as the complex number $V=V_\text{rms}\,e^{j\varphi}$ obtained by freezing the frequency $\omega_0$.  This complex number is called `Fresnel vector', `phase vector', or `phasor'.  When $V_\text{rms}$ and $\varphi$ are allowed to vary (slowly), the term `analytic signal' is used instead.  The object of our interest is the primary measurement of the phase $\varphi$, focusing on small randomly fluctuating angles.

\begin{figure}
\centering\namedgraphics{0.8}{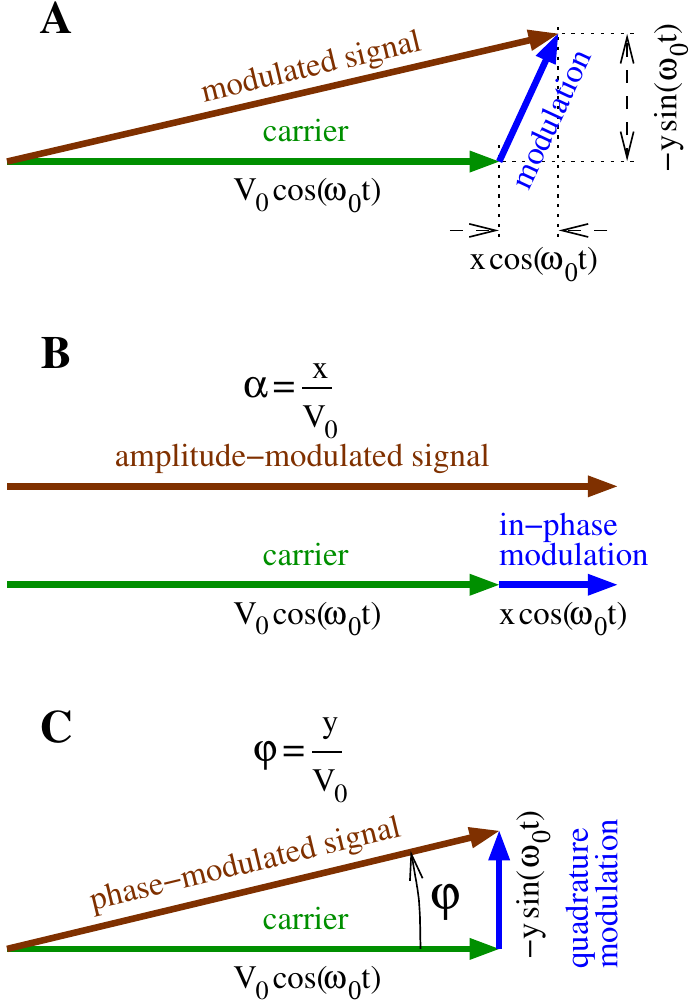}{\textwidth}
\caption{Phasor representation of a randomly modulated signal.}
\label{fig:cal-modulated}
\end{figure}
Using the peak amplitudes instead of the rms amplitudes, we introduce the signal (Fig.~\ref{fig:cal-modulated}\,A)
\begin{align}
v_i(t)=V_0[1+\alpha(t)]\cos[\omega_0t+\varphi(t)]~, 
\label{eqn:cal-noisy-sinusoid-polar}
\end{align}
which defines the random fractional amplitude $\alpha(t)$ and the random phase $\varphi(t)$.  
The above can be rewritten as \cite{rice44bstj,rice45bstj} 
\begin{align}
v_i(t)=V_0\cos\omega_0t + x(t)\cos\omega_0t - y(t)\cos\omega_0t~,
\label{eqn:cal-noisy-sinusoid-cartesian}
\end{align}
where $x(t)$ and $y(t)$ random processes.  The two above representations of $v_i(t)$ are connected by
\begin{align}
\label{eqn:cal-alpha-exact}
\alpha(t)
&= V_0 \sqrt{\Bigl[1+\frac{x}{V_0}\Bigr]^2+\Bigl[\frac{y}{V_0}\Bigr]^2}-1
&&\text{fractional amplitude}\\[1ex]
\label{eqn:cal-phi-exact}
\varphi(t)&= \arg(y,V_0+x)~.
&&\text{phase, $\varphi\in(-\pi,\pi]$}
\end{align}
In low-noise conditions, that is, $|x/V_0|\ll1$ and $|y/V_0|\ll1$, $\alpha(t)$ and $\varphi(t)$ turn into
\begin{align}
\label{eqn:cal-alpha-phi}
\alpha(t)=\frac{x}{V_0}	\quad\text{and}\quad
\varphi(t)=\frac{y}{V_0}~.
\end{align}
For calibration purposes, we need a signal with pure AM ($\varphi=0$, thus $y=0$), and one with pure PM ($\alpha=0$, thus $x=0$), as in Fig.~\ref{fig:cal-modulated}\,B-C\@.

Phase noise is usually described in term of $S_\varphi(f)$, namely, the one-sided power spectral density (PSD) of $\varphi(t)$ as a function of the Fourier frequency $f$. The physical dimension of $S_\varphi(f)$ is \unit{rad^2\,Hz^{-1}}.  Similarly, $S_\alpha(f)$ is the PSD of $\alpha(t)$.  $S_\alpha(f)$ has the dimension of \unit{Hz^{-1}}.  The technical quantity $\mathcal{L}(f)$ is also used, defined as $\mathcal{L}(f)=\frac12S_\varphi(f)$ and given in dBc (dB carrier).  The technical unit dBc is also used for AM noise.
Trusting the stationarity and the ergodicity of the random process, that is, the repeatability and the reproducibility of the experiment, the PSD is measured as the average square modulus of the one-sided Fourier transform normalized for the power-type signals.  The general background on phase noise and on frequency stability is available from numerous references, among which we prefer  \cite{chronos:frequency,kroupa:frequency-stability,ieee99std1139,ccir90rep580-3}.

\subsection{State of the art} 
Generally, the measurement of AM and PM noise is done with a suitable detector followed by a fast Fourier transform (FFT) analyzer.
In most practical cases, the phase detector is a double balanced mixer \cite{nelson04fcs,agilent:E5500-phase-noise,aeroflex:PN9000-phase-noise} with the two inputs in quadrature.  The amplitude detector \cite{rubiola05arxiv-am-noise} is a power detector, that is, a diode used in the quadratic region, or a double balanced mixer with the two inputs in phase.  
A balanced bridge (often referred to as `interferometer') with amplification and synchronous detection of the noise sidebands is used when the highest sensitivity is required \cite{sann68mtt,labaar82microw}.  The sensitivity is limited by the equivalent temperature of the instrument \cite{ivanov98uffc}.  Improved sensitivity is obtained by correlation and averaging, with two separate---thus independent---systems that measure the same device under test (DUT) \cite{vessot64nasa,walls76fcs}.  The dual-bridge with correlation exhibits the highest reported sensitivity, limited by the thermal uniformity of the instrument instead of the absolute temperature \cite{rubiola2000rsi-correlation}.
The bridge works well for the measurement of two-port components, yet it is difficult to use for the oscillator PM noise, and virtually unsuitable to the oscillator AM noise.

\begin{figure}[t]
\centering\namedgraphics{0.8}{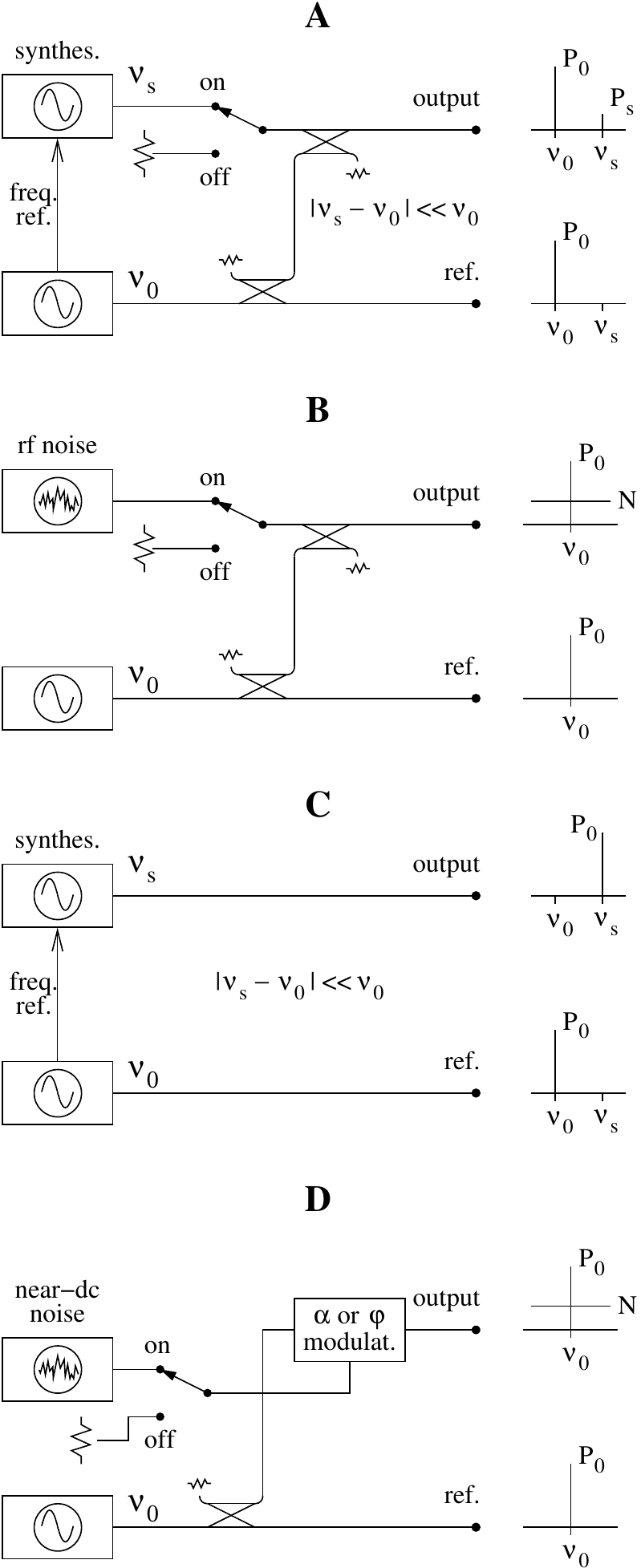}{\textwidth}
\caption{Generally used calibration methods.}
\label{fig:cal-methods}
\end{figure}

Calibration is the accurate measurement of the amplitude-to-voltage gain $k_\alpha$ and of the phase-to-voltage gain $k_\varphi$ of the detector. 
The calibration relies upon a reference of phase modulation and of amplitude modulation.  A deterministic modulation is often used because the uncertainty of the base-band instruments that follow the detector can be made small as compared to that of the RF/microwave section.  Figure~\ref{fig:cal-methods} shows the calibration schemes generally used.

Following the scheme A, a sideband of frequency $\nu_s$ close to $\nu_0$ is added to the carrier in one arm of the phase detector.  This is equivalent to phase modulation plus amplitude modulation of rms value $\sqrt{P_s/2P_0}$, and frequency $\nu_b=|\nu_s-\nu_0|$. $P_s$ and $P_0$ are the sideband power and the carrier power, respectively.
The scheme B is similar to the scheme A, but for the use of RF/microwave random noise instead of the sideband.  The close-in noise is $S_\alpha(f)=S_\varphi(f)=N/P_0$, where $N$ is the noise PSD\@.

The schemes A and B suffer from the impossibility to divide AM from PM\@.  Instead, an equal amount AM and PM is present.  Hence, a standard based on Fig.~\ref{fig:cal-methods}\,A-B should be referred to as a standard of \emph{SNR} (signal-to-noise-ratio), rather than of AM and PM noise.  The problem is that the mixer used as the phase detector is not perfectly balanced, for the dc offset is affected by the input power.  Hence, it turns the AM noise into near-dc noise, which is mistaken for PM noise.  The AM noise rejection can be of 15--35 dB, depending on the operating power, frequency and quadrature error, on the mixer type, and on technology \cite{rubiola06arxiv-am-to-pm-pollution,rubiola07uffc-am-to-pm-pollution}.  Using the mixer at a sweet point of zero-sensitivity to AM noise, as suggested in \cite{brendel75im,cibiel02uffc}.  Consequently, the \emph{use} of this standard in the calibration of PM noise measurements based on the saturated mixer is incorrect.  

In the scheme of Fig.~\ref{fig:cal-methods}\,C, two sinusoidal signals are sent to the mixer.  These signals have frequency $\nu_0$ and $\nu_s$, with a small difference $\nu_b=|\nu_s-\nu_0|$, and the same power of the final application.  The detected signal is a sinusoid of frequency $\nu_b$ crossing the 0 V dc axis with slope $\frac{dv}{dt}=\omega_bk_\varphi$, from which $k_\varphi$ can be measured using a digital oscilloscope.
Of course, this scheme can only be used only for the calibration of PM measurements, not for AM\@.  The major problem with this approach is that it is impossible to assess the mixer's unwanted sensitivity to AM noise.

The scheme of Fig.~\ref{fig:cal-methods}\,D makes use of a true amplitude or phase modulator.  This is definitely correct, provided the modulator was calibrated.  Avoiding AM to PM crosstalk, that is, avoiding residual AM in the phase modulator and residual PM in the amplitude modulator, is the critical point.    

Presently, the state of the art in AM and PM noise standards is the NIST secondary standard of AM and PM noise \cite{walls93im,nist-calibration}.
This standard is based on the scheme of Fig.~\ref{fig:cal-methods}\,B, thus it makes the use of the mixer incorrect as discussed before.  Additionally, however accurate, this standard is a secondary one.  Commercial standards of this type also exist, like that of Techtrol Cyclonetics Inc.\ (TCI) \cite{tci-ant-calibration}.

A reference modulator (Fig.~\ref{fig:cal-methods}\,D) is in use at NIST\footnote{Craig Nelson, Phase Noise Measurements, 2006 FCS tutorial}, calibrated with the secondary standard.
Yet, nothing is said about how to avoid the AM to PM crosstalk.

In the BIPM web site\footnote{http://kcdb.bipm.org/appendixC/\ldots}, only the French LNE reports on the accuracy of phase noise measurements.
The uncertainty is of 2 dB\@.  However disappointing, this value accounts for all conditions of routine measurements.
The NIST\footnote{http://ts.nist.gov/MeasurementServices/Calibrations/oscillators.cfm} reports about typical accuracy of 1 dB in both AM and PM noise measurements.
An old article on the bridge \cite{fikart72mtt-noise-accuracy}.
The declared uncertainty is of 1.9 dB at the NPL, UK.

\section{Method}
This article proposes a method for the primary calibration of AM and PM measurements, based on the following ideas.
\vspace{-1ex}
\begin{itemize}\addtolength{\itemsep}{-1.2ex}
\item We first understand that the calibration of the detector (amplitude or phase) is the critical point, while the accuracy of the near-dc complementary instruments (FFT, etc.) can be easily made negligible for all practical purposes. 

\item A reference modulator is used to calibrate the gain of the detector.  Of course, we need an amplitude modulator free from phase 
modulation, and a phase modulator free from amplitude modulation.
The conceptual scheme of the modulator is shown in Fig.~\ref{fig:cal-modulator}.  The amplitude modulation is obtained by adding to the carrier $V_0\cos(\omega_0t)$ a signal $x(t)\cos(\omega_0t)$.  
Thus, the amplitude modulation is $\alpha(t)=x(t)/V_0$ [Equations \req{eqn:cal-noisy-sinusoid-polar} and \req{eqn:cal-noisy-sinusoid-cartesian}].  Similarly, the phase modulation is obtained by adding a signal $-y(t)\cos(\omega_0t)$, so that $\varphi(t)=y(t)/V_0$
\begin{figure}[t]
\centering\namedgraphics{0.8}{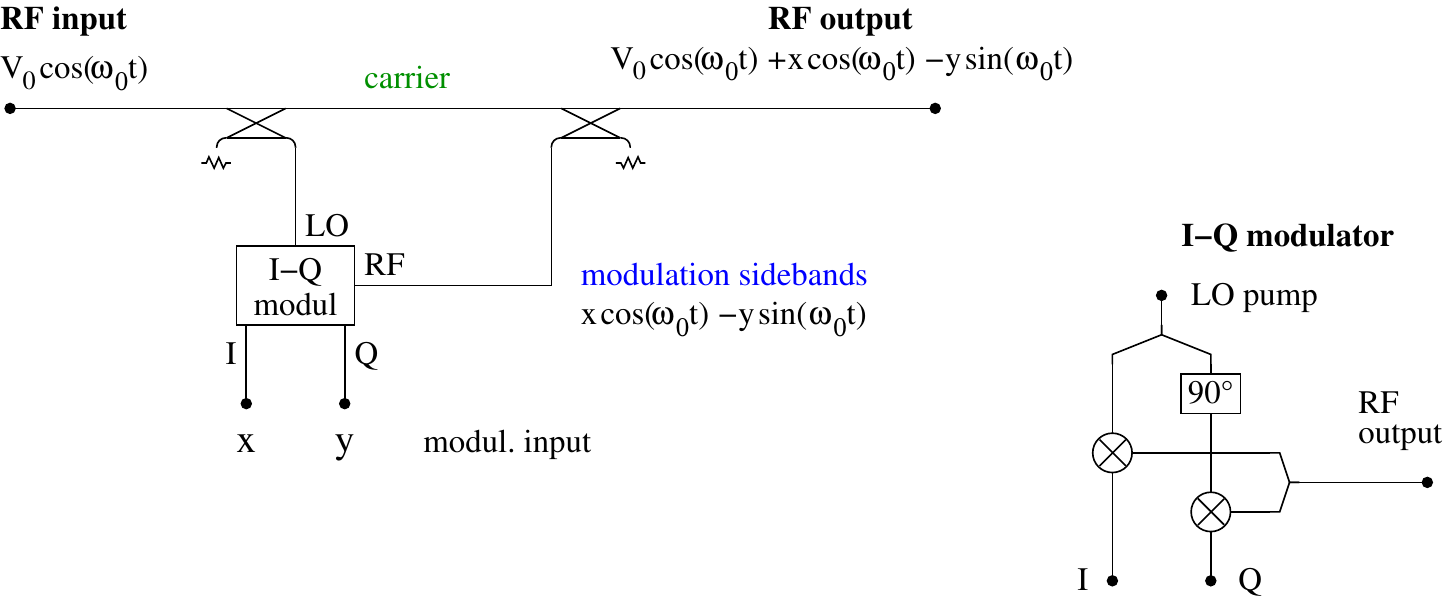}{\textwidth}
\caption{Reference modulator.}
\label{fig:cal-modulator}
\end{figure}

\item For our purpose, it is necessary to fix the quadrature error $\psi$ and the gain asymmetry $\epsilon$ of the real I-Q modulators and detectors (Fig.~\ref{fig:cal-iq-ware}).  This can be done with null and differential measurements only.

\item Real phase detectors show a residual sensitivity to amplitude modulation.  This is due to non-perfect saturation.  Conversely, power detectors are not sensitive to phase modulation, provided the video\footnote{The output is usually called `video.'} bandwidth is large enough to prevent the discriminator effect due to the internal memory.  Consequently, we can use null measurement with a power detector to find the phase $\theta$ of the pump.  $\theta$ is an arbitrary quantity resulting from the electrical layout, to be corrected in order to set the appropriate phase relationship between the modulation sidebands and the carrier.

\item Switching on and off the modulation sideband, the modulation is measured as a power ratio.  Thus, $\alpha_\text{rms}=\sqrt{P_x/P_0}$, and $\varphi_\text{rms}=\sqrt{P_y/P_0}$, where $P_0$, $P_x$ and $P_y$ are the power of the carrier, of the in-phase modulation, and of the quadrature modulation.   

\end{itemize}
It is to be remarked that the entire procedure proposed here relies only on null, differential, and ratiometric measurements.  
This will be discussed in detail after explaining some technical tools.

\section{Mixers, scalar products, and Hilbert spaces}\label{sec:scalar-product}
It is well known in the classical theory of telecommunications \cite{wonzecraft:communication,viterbi:communication,lindsey:telecomm} that with the appropriate definition of the scalar product, the space of the sinusoids at a given frequency ($\omega_0$) has the structure of an Hilbert space.  In fact
\begin{enumerate}
\item the space is a inner-product space, that is, the scalar product $(x,y)$ has the usual properties of conjugation and linearity, and $(x,x)\ge0$, where $(x,x)=0$ only if $x=0$,
\item the norm is $||x||=\sqrt{(x,x)}$,
\item the space is complete.
\end{enumerate}
In such space, we can define a orthogonal unitary base, so that all signals can be expressed as a linear combination of the base.
 
\subsection{Complex signals}
The scalar product between two complex signals $s(t)$ and $r(t)$ is defined as
\begin{align}
(s,r) &= \frac1T\int_0^T s(t)\,r^*(t)\:dt
\label{eqn:cal-scalar-complex}
\end{align}
where $T$ is a suitable integration time, and the superscript * denotes the complex conjugate.
The base function is the complex sinusoid
\begin{align}
&r(t) =  e^{j\omega_0t}		&&\text{reference} \label{eqn:cal-reference-complex}\\
&||r(t)|| = 1
\end{align}
Mathematically, the integration time should be $T=n\smash{\frac{2\pi}{\omega_0}}$, integer $n$, so that the $\omega_0$ and $2\omega_0$ terms are canceled.  In practice, a sufficiently long $T\gg\frac{2\pi}{\omega_0}$ makes the $2\omega_0$ terms negligible for any practical purposes. 

Let us consider the scalar product of the signal
\begin{align}
s(t) = Ve^{j\omega_0t+\varphi}	&&\text{signal} \label{eqn:cal-signal-complex}
\end{align}
on the reference $r(t)$.
Expanding Eq.~\req{eqn:cal-scalar-complex}, the scalar product $v=(s,r)$ is 
\begin{align}
\begin{split}
v &=\frac1T\int_0^T V
	\Bigl[\exp(j\omega_0t+\varphi) \exp(-j\omega_0t)\Bigr]\,dt \\[1ex]
&=\frac1T\int_0^T V
	\Bigl[\cos(\omega_0t+\varphi)+j\sin(\omega_0t+\varphi)\Bigr]
	\Bigl[\cos(\omega_0t)-j\sin(\omega_0t)\Bigr]\,dt \\[1ex]
&=\frac1T\int_0^T V \Bigl\{
	\Bigl[\cos(\omega_0t)\cos(\varphi)-\sin(\omega_0t)\sin(\varphi)\Bigr] +{}\\[-1ex] 
	&\hspace{14ex}
	+j\Bigl[\cos(\omega_0t)\sin(\varphi)-\sin(\omega_0t)\cos(\varphi)\Bigr]
	\Bigr\} \\[0ex]
	&\hspace{32ex}\times
	\Bigl[\cos(\omega_0t)-j\sin(\omega_0t)\Bigr]\,dt
\end{split}\nonumber
\intertext{and finally}
v &= V\bigl[\cos(\varphi)+j\sin(\varphi)\bigr] 
	\qquad\qquad\text{(after removing the $2\omega_0$ terms)}\\[1ex]
   &= Ve^{j\varphi}
\end{align}
Note that the signal $s(t)$ can be expressed as
\begin{align}
s(t) &= v\,r(t)
\end{align}
The important fact is that the scalar product freezes the $\omega_0$ carrier frequency, thus it gives the phasor $v=Ve^{j\varphi}$ associated to $s(t)$.  $V$ is the rms voltage.

\subsection{Real signals}
Dealing with real signals $s(t)$ and $r(t)$, the scalar product \req{eqn:cal-scalar-complex} turns into
\begin{align}
(s,r) &= \frac1T\int_0^T s(t)\,r(t)\:dt
\label{eqn:cal-scalar-real}
\end{align}
because $r(t)=r^*(t)$, thus the complex conjugate is no longer needed.
A two-dimension base is necessary to replace the complex sinusoid
\begin{align}
r_I(t) &=+\sqrt{2}\cos(\omega_0t)	&&\text{in-phase reference}
	\label{eqn:cal-reference-I-real}\\[1ex]
r_Q(t)&=-\sqrt{2}\sin(\omega_0t)	&&\text{quadrature reference}
	\label{eqn:cal-reference-Q-real}
\end{align}
The factor $\sqrt{2}$ is necessary for $||r_I||=1$ and $||r_q||=1$.
Let us consider the scalar product of
\begin{align}
s(t)	&=V\sqrt{2}\cos(\omega_0t+\varphi)	&&\text{signal}
	\label{eqn:cal-signal-real}
\end{align}
on the base functions.
Expanding Eq.~\req{eqn:cal-scalar-real}, the scalar product $v_I=(s,r_I)$ is 
\begin{align}
\begin{split}
v_I 
&=\frac1T\int_0^T 2V\cos(\omega_0t+\varphi)\cos(\omega_0t)\,dt\\
&=\frac1T\int_0^T 
	2V\Bigl[\cos(\omega_0t)\cos(\varphi)-\sin(\omega_0t)\sin(\varphi)\Bigr]
	\cos(\omega_0t)\,dt \\
&=\frac1T\int_0^T 
	2V\Bigl[\cos^2(\omega_0t)\cos(\varphi)-
	\sin(\omega_0t)\cos(\omega_0t)\cos(\varphi)\Bigr]\,dt \\[1ex]
&=  V\cos(\varphi)
	\qquad\qquad\text{(after removing the $2\omega_0$ terms)}
\end{split}
\label{eqn:cal-scalar-I-real}
\intertext{Similarly, the scalar product $v_Q=(s,r_Q)$ is}
\begin{split}
v_Q
&=\frac1T\int_0^T 2V\cos(\omega_0t+\varphi)\,[-\sin(\omega_0t)]\,dt\\
&=\frac1T\int_0^T 
	2V\Bigl[\cos(\omega_0t)\cos(\varphi)-\sin(\omega_0t)\sin(\varphi)\Bigr]
	\Bigl[-\sin(\omega_0t)\Bigr]\,dt \\
&=\frac1T\int_0^T 
	2V\Bigl[\sin(\omega_0t)\cos(\omega_0t)\cos(\varphi)
	+ \sin^2(\omega_0t)\sin(\varphi)\Bigr]\,dt \\[1ex]
&=  V\sin(\varphi)
	\qquad\qquad\text{(after removing the $2\omega_0$ terms)}
\end{split}
\label{eqn:cal-scalar-Q-real}
\end{align}
Note that the real signal $s(t)$ can be expressed as 
\begin{align}
s(t) &= v_I\,r_I(t) + v_Q\,r_Q(t)~.
\end{align}
Once again, the signal $s(t)$ is completely defined by the scalar projections on the base functions.  The proof is trivial
\begin{align}
s(t) &= V\cos(\varphi) \, \sqrt{2}\cos(\omega_0t) -
		V\sin(\varphi) \, \sqrt{2}\sin(\omega_0t)\\
& = V\sqrt{2}\cos(\omega_0t+\varphi)~.
\end{align}

\begin{figure}[t]
\centering\namedgraphics{0.8}{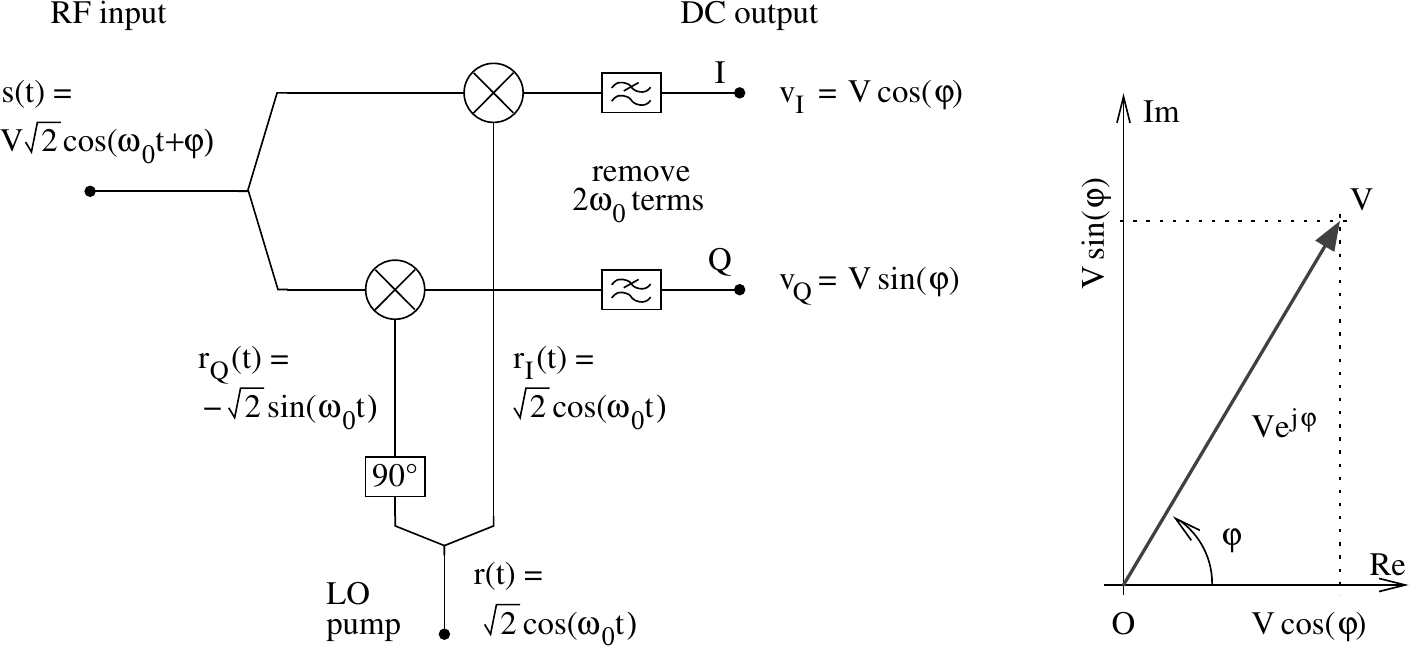}{\textwidth}\\[0ex]
\caption{Ideal I-Q detector.}
\label{fig:cal-iq-detect}
\vspace{2em}
\centering\namedgraphics{0.8}{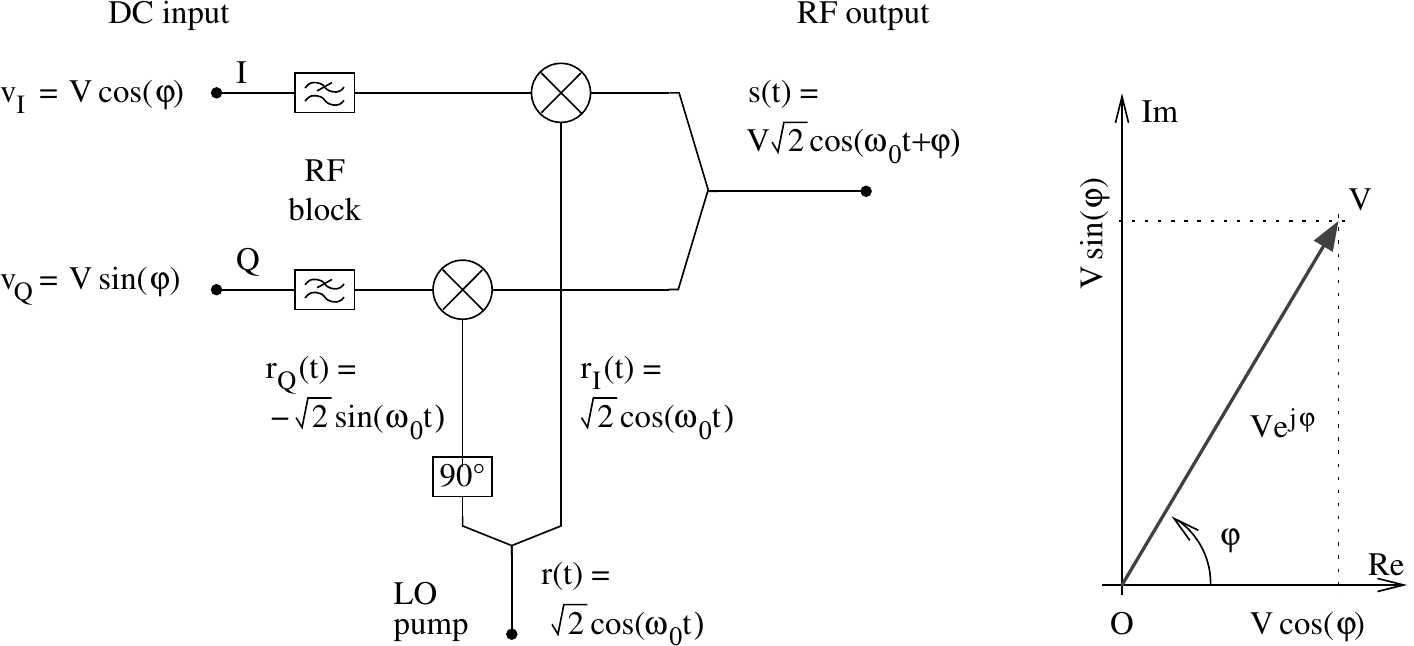}{\textwidth}
\caption{Ideal I-Q modulator.}
\label{fig:cal-iq-modul}
\end{figure}
\subsection{I-Q detector and modulator}
The I-Q detector is a hardware implementation of Eq.~\req{eqn:cal-scalar-I-real} and \req{eqn:cal-scalar-Q-real}.  Fig.~\ref{fig:cal-iq-detect} shows the conceptual scheme of an ideal I-Q detector, based on analog multipliers.  The rules of energy management and impedance matching are still not considered here.  In the geometrical analogy, the I-Q detector extracts the real and imaginary part of the input phasor.

The I-Q modulator (Fig.~\ref{fig:cal-iq-detect}) is similar to the detector, but the signal flow is reversed.  In the geometrical analogy, the I-Q modulator turns $v_I$ into a real phasor and $v_Q$ into an imaginary phasor, and builds the output phasor by adding the real part and imaginary part.

\begin{figure}[t]
\centering\namedgraphics{0.8}{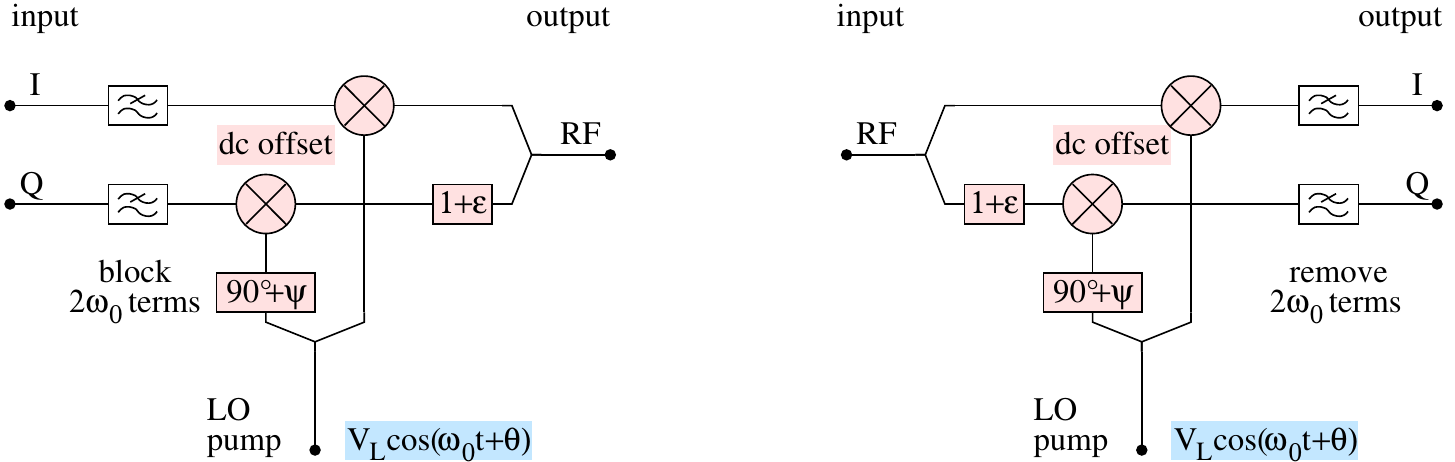}{\textwidth}
\caption{I-Q modulator and detector.}
\label{fig:cal-iq-ware}
\vspace{2em}
\centering\namedgraphics{0.8}{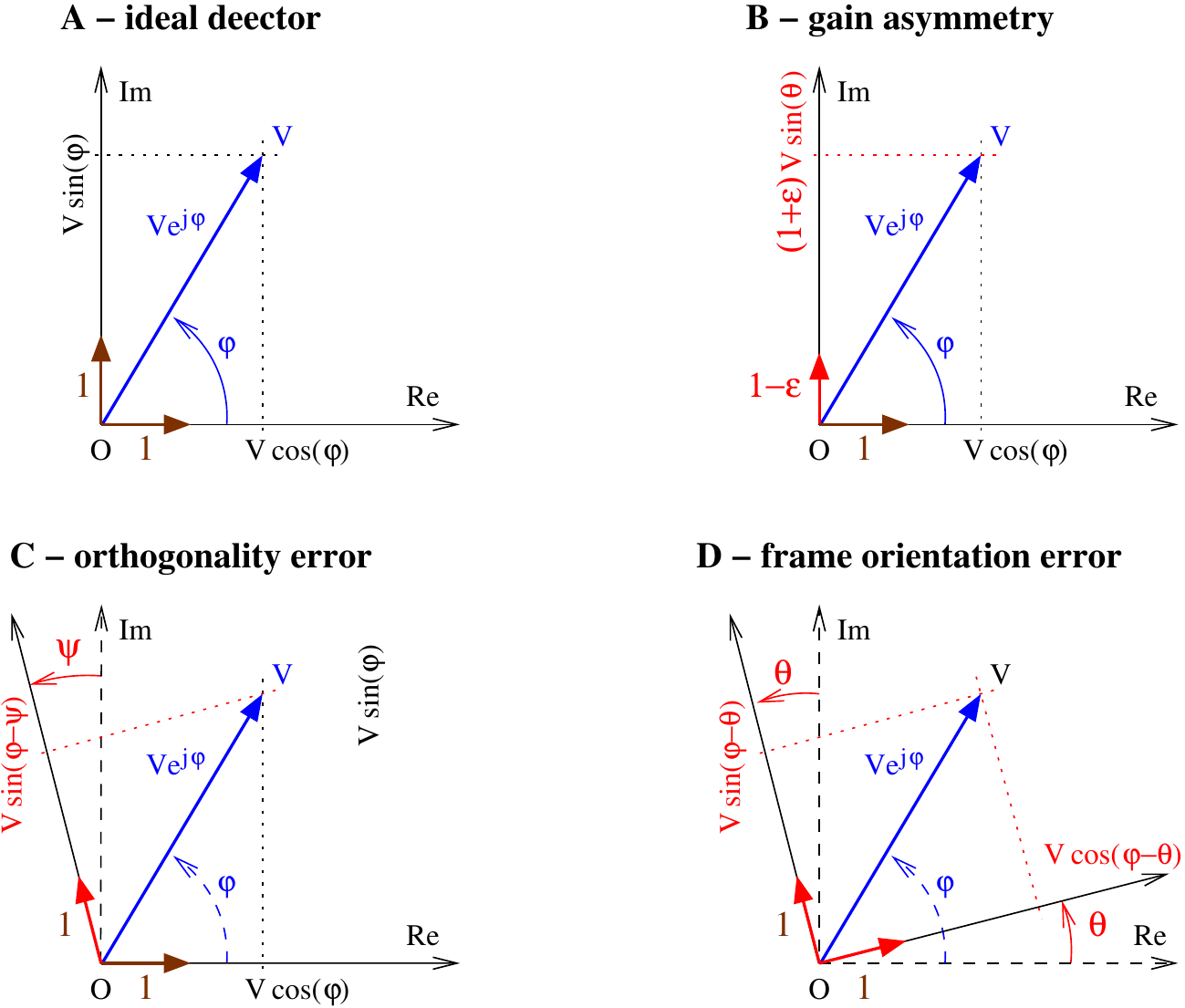}{\textwidth}
\caption{Errors of the real I-Q detector.}
\label{fig:cal-iq-errors}
\end{figure}
\subsection{Defects of the real I-Q detectors and modulators}
Practical I-Q detectors and modulators are based on double balanced mixers and on baluns.  See 
the reference \cite{rubiola06arxiv-tutorial-on-mixers} for a tutorial.
Figures~\ref{fig:cal-iq-ware} and \ref{fig:cal-iq-errors} show the defects of a practical I-Q detector and modulator, detailed underneath.
In addition, we stress that the inputs can not be driven with a voltage source.
\begin{description}
\item[DC offset and LO crosstalk.] Due to the asymmetry of the diode ring or of the baluns, a dc offset appears at the output of the detector.  The same phenomenon manifests as the imperfect suppression of the carrier at the modulator output when the inputs are set to zero.

\item[Loss.] The mixer operates with the LO input saturated.  Accordingly, the peak amplitude of the reference signals takes the value $V_L$, which generally differs from $\sqrt{2}$.  For a given mixer used as a detector, $V_L$ can be calculated from the mixer loss $\ell$ using $V_L=\frac{2\,\text{V}}{\ell}$.  The typical loss of an I-Q detector is $\ell=2.8$ (9 dB, i.e., the 6 dB of a mixer plus the 3 dB of the input power splitter), we calculate $V_L=700$ mV\@.

\item[Gain asymmetry.] The mixer loss is not the same for the I and Q arm.  A difference of 5\% is common.

\item[Orthogonality error.] The phase shift of the pump in the Q arm deviates by an angle $\psi$ from the quadrature.  This error, which can be of a few degrees, is frequency dependent.

\item[Frame orientation error.] The arbitrary length of the cable connecting the LO pump to the reference turns into an arbitrary phase $\theta$, which affects both the I and Q arms.  
\end{description}
Correcting these defect is a central point in our work.

\section{Correction of the I-Q detector}\label{cal:sec-fix-iqdet}
\begin{figure}[t]
\centering\namedgraphics{0.8}{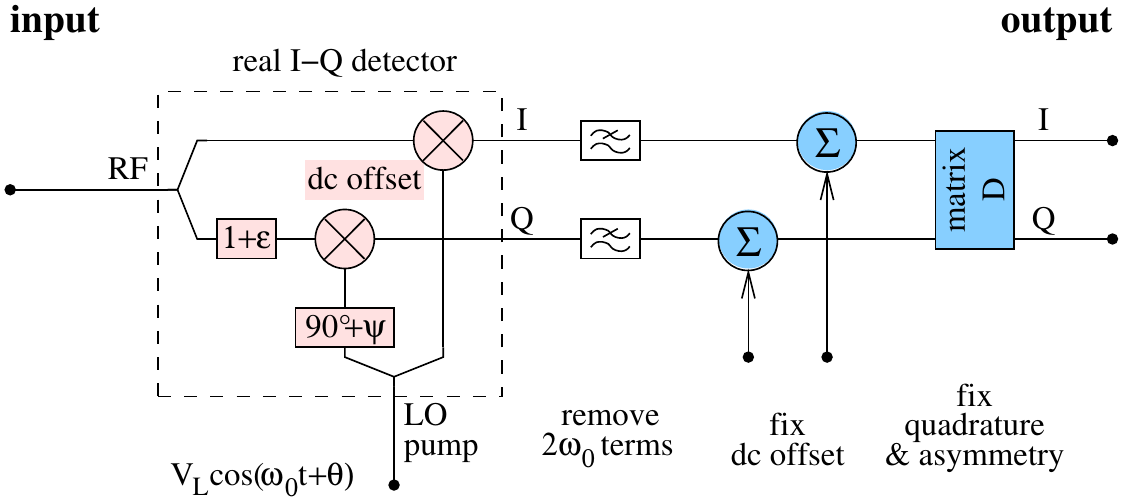}{\textwidth}
\caption{Turning a real I-Q detector into a nearly ideal one.}
\label{fig:cal-detect-correct-b}
\end{figure}

This section describes how to compensate the defects of a I-Q detector, turning it into a virtually ideal one, as shown in Fig.~\ref{fig:cal-detect-correct-b}.  The offset is removed by adding an appropriate dc. The the orthogonality defect $\psi$ and the gain asymmetry $\epsilon$ are fixed by taking a linear combination of the two outputs.  The phase of the LO pump is still let arbitrary, as it results from the electrical layout. 

\subsection{Measurement of the errors \boldmath$\psi$ (orthogonality) and \boldmath$\epsilon$ (symmetry)}
\begin{figure}[t]
\centering\namedgraphics{0.75}{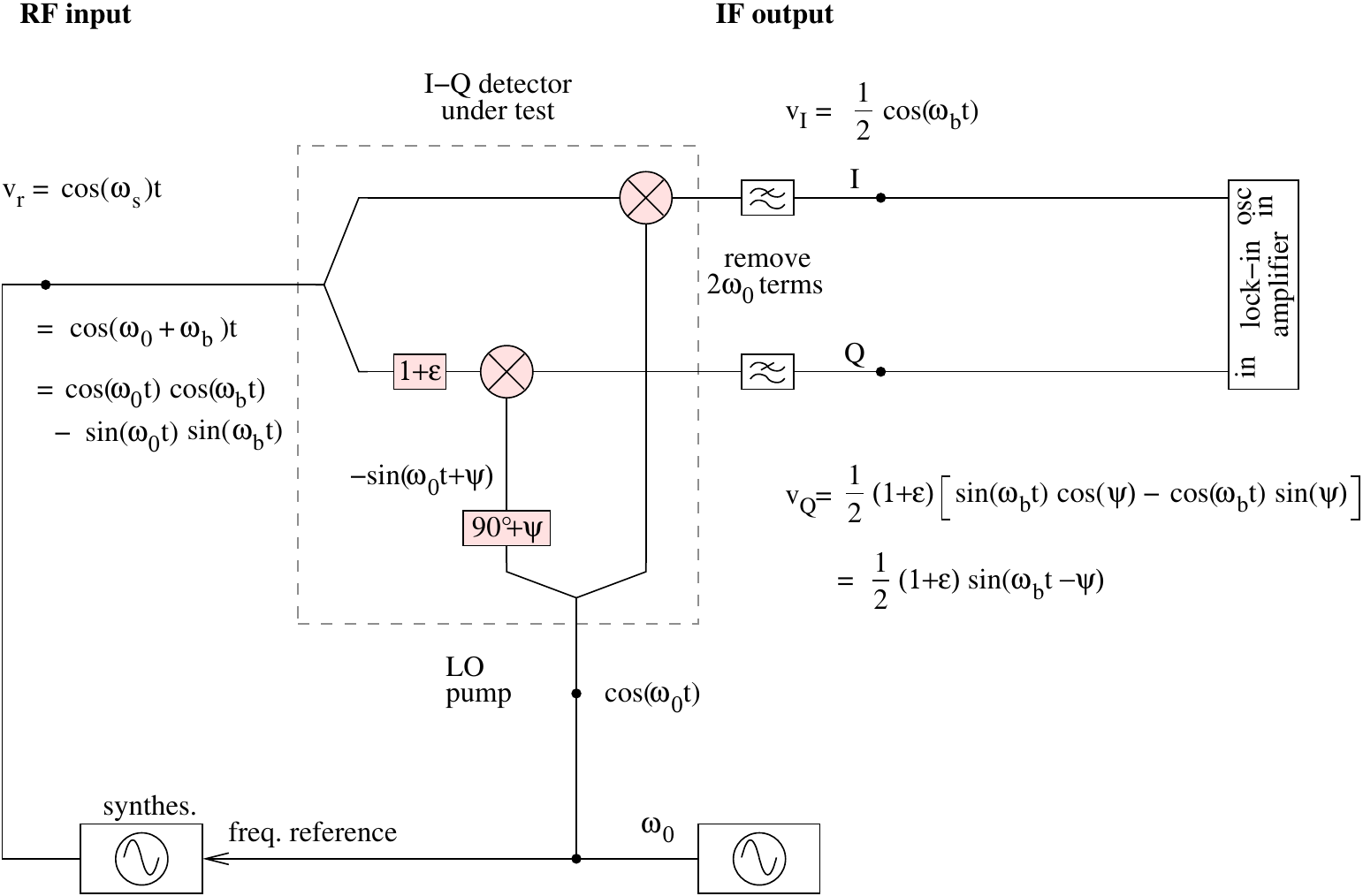}{\textwidth}
\caption{Measurement of the orthogonality and symmetry defect of a real I-Q detector.}
\label{fig:cal-detect-correct-a}
\end{figure}

First, the dc offset is compensated for after inspection with a dc voltmeter, with the RF input terminated.  Afterwards, the matrix $D$ corrects for $\psi$ and $\epsilon$.  The method to obtain the elements of $D$ from the measurements is shown in Figures~\ref{fig:cal-detect-correct-a}, and described underneath.

We feed a low-power sideband
\begin{align}
v_{r}(t) &= \cos(\omega_0+\omega_b)t\\[0.5ex]
&= \cos(\omega_0t)\cos(\omega_bt)-\sin(\omega_0t)\sin(\omega_bt)
\end{align}
in the RF input of the detector.  For the sake of simplicity, the amplitude coefficients are omitted.  The frequency $\omega_s=\omega_b+\omega_0$ is chosen for $\omega_b$ to be an appropriate audio frequency (1--2 kHz).  The LO input is saturated by a signal $\cos(\omega_0t)$ of suitable power.  The output signals 
\begin{align}
v_{I}(t) &= \frac12 \cos(\omega_bt)\\[1ex]
v_{Q}(t) 
&= \frac12 (1+\epsilon)\bigl[\sin(\omega_bt)\cos(\psi)-\cos(\omega_bt)\sin(\psi)\bigr]\\[0.5ex]
&= \frac12 (1+\epsilon)\sin(\omega_bt-\psi)
\end{align}
are measured with a lock-in amplifier, which gives $\psi$ and $\epsilon$.  The lock-in takes the $I$ signal as the oscillator reference.  Most lock-in amplifiers measure only the amplitude of the input, not the amplitude of the oscillator reference.  If so, input and oscillator reference are to be interchanged in order to get both amplitudes.  Interchanging the lock-in inputs also helps to get an accurate measure of $\psi$.

\subsection{Correction of the orthogonality and symmetry errors}\label{ssec:cal-fix-detector}
The detector is corrected by introducing the matrix $D$ at the output (Fig.~\ref{fig:cal-detect-correct-b}).
In the remaining part of this section we denote with $v'_I(t)$ and $v'_Q(t)$ the outputs of the I-Q detector, and with $v_I(t)$ and $v_Q(t)$ the outputs of the \emph{corrected} detector, after the matrix $D$.  Thus, 
\begin{align}
\begin{bmatrix}v_I\\v_Q\end{bmatrix} &=
\begin{bmatrix}d_{11}&d_{12}\\d_{21}&d_{22}\end{bmatrix}
\begin{bmatrix}v'_I\\v'_Q\end{bmatrix}
\end{align}
The elements of $D$ are calculated by separating the $\cos(\omega_0t)$ and $\sin(\omega_0t)$ parts of the input signal, i.e., the real and imaginary parts of the input phasor.  Thus, when the RF input is 
\begin{align}
v_r(t)=\cos(\omega_0t)\cos(\omega_bt) \qquad\text{(real part)}
\end{align}
the output
\begin{align}
v'_I(t)	&=\frac12 \cos(\omega_bt)\\
v'_Q(t)	&=-\frac12 (1+\epsilon)\cos(\omega_bt)\sin\psi
\intertext{must be transformed into}
v_I(t)	&=\frac12 \cos(\omega_bt)\\
v_Q(t)	&=0~.
\end{align}
This is equivalent to state that 
\begin{align}
\begin{bmatrix}d_{11}&d_{12}\\d_{21}&d_{22}\end{bmatrix}
\begin{bmatrix}\frac12 \cos(\omega_bt)\\
	-\frac12 (1+\epsilon)\cos(\omega_bt)\sin\psi\end{bmatrix}=
\begin{bmatrix}\cos(\omega_bt)\\0\end{bmatrix}
\end{align}
holds for all $t$, which yields  
\begin{gather}
d_{11}=2\\
d_{12}=0\\
d_{21}-d_{22}(1+\epsilon)\sin\psi = 0~.
\end{gather}
Similarly, when the RF input is 
\begin{align}
v_r(t)=-\sin(\omega_0t)\sin(\omega_bt) \qquad\text{(imaginary part)}
\end{align}
the output
\begin{align}
v'_I(t)	&=0\\
v'_Q(t)	&=\frac12 (1+\epsilon)\sin(\omega_bt)\cos\psi
\intertext{must be transformed into}
v_I(t)	&=0\\
v_Q(t)	&=\frac12 \sin(\omega_bt)~.
\end{align}
This is equivalent to state that 
\begin{align}
\begin{bmatrix}d_{11}&d_{12}\\d_{21}&d_{22}\end{bmatrix}
\begin{bmatrix}0\\
	\frac12 (1+\epsilon)\sin(\omega_bt)\cos\psi\end{bmatrix}=
\begin{bmatrix}0\\\sin(\omega_bt)\end{bmatrix}
\end{align}
holds for all $t$, which yields  
\begin{align}
d_{22} = \frac{2}{(1+\epsilon)\cos\psi}~.
\end{align}
Finally, combining the above results we get
\begin{align}
D &= 2\begin{bmatrix}1&0\\[1ex]\displaystyle\frac{\sin\psi}{\cos\psi}
	&\displaystyle\frac{1}{(1+\epsilon)\cos\psi}\end{bmatrix}
\end{align}

\subsection{Correction of the pump phase \boldmath$\theta$}
\begin{figure}[t]
\centering\namedgraphics{0.8}{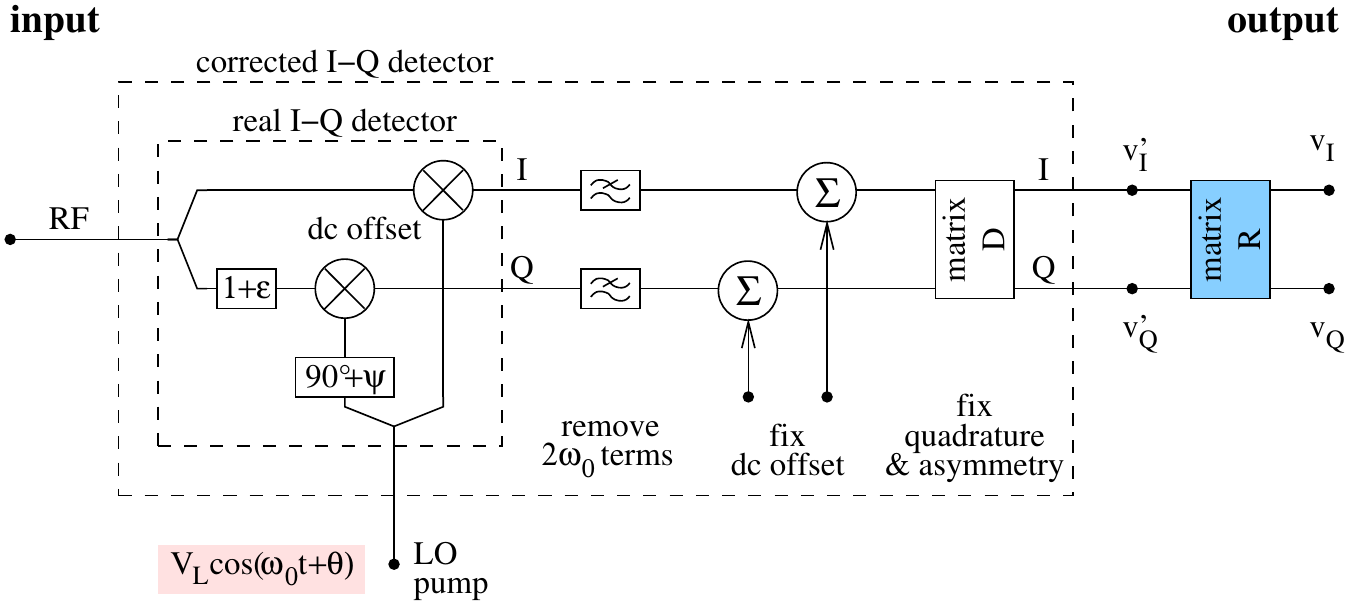}{\textwidth}
\caption{Correction of the arbitrary phase $\theta$ of the LO pump.}
\label{fig:cal-detect-correct-c}
\end{figure}

As a result of the previous sections, the I-Q detector is corrected for the amplitude asymmetry $\epsilon$ and for the quadrature error $\psi$.  The phase $\theta$ of the LO pump is still arbitrary.  This can be corrected (Fig.~\ref{fig:cal-detect-correct-c}) by introducing at the matrix 
\begin{align}
\label{eqn:cal-det-fix-theta-r}
R &= \begin{bmatrix}\cos\theta&-\sin\theta\\\sin\theta&\cos\theta\end{bmatrix}~,
\end{align}
which rotates the output by $\theta$.
This requires an absolute reference of angle, which we still have not.  For the sake of completeness, we show the mathematical derivation of the matrix $R$.

We start by feeding the signal
\begin{align}
v_r(t)=x\cos(\omega_0t)-y\sin(\omega_0t)
\end{align}
at the RF input.  The I-Q internal pump signals are $\cos(\omega_0t+\theta)$ and $\cos(\omega_0t+\theta)$.  As before, we denote with $v'_I(t)$ and $v'_Q(t)$ the outputs of the I-Q detector (now corrected by the matrix $D$), and with $v_I(t)$ and $v_Q(t)$ the outputs of the \emph{corrected} detector, after the matrix $R$. 
In this section we discard the factor 2 introduced by the matrix $D$, as we had an ideal I-Q detector without correction.  Thus, the output signals are
\begin{align}
v'_I(t)&=\bigl[x\cos(\omega_0t)-y\sin(\omega_0t)\bigr]\cos(\omega_0t+\theta)\\
	&=\bigl[x\cos(\omega_0t)-y\sin(\omega_0t)\bigr]
		\bigl[\cos(\omega_0t)\cos\theta-\sin(\omega_0t)\sin\theta\bigr]\\
	&=\frac12 \bigl[x\cos(\theta)+y\sin\theta\bigr]
	\qquad\text{(after removing the $2\omega_0$ terms)}\\[2ex]
v'_Q(t)&=\bigl[x\cos(\omega_0t)-y\sin(\omega_0t)\bigr]
		\bigl\{-\sin(\omega_0t+\theta)\bigr\}\\
	&=\bigl[x\cos(\omega_0t)-y\sin(\omega_0t)\bigr]
		\bigl\{-\bigl[\sin(\omega_0t)\cos\theta+\cos(\omega_0t)\sin\theta\bigr]\bigr\}\\
	&=-\frac12 \bigl[x\sin(\theta)-y\cos\theta\bigr]
	\qquad\text{(after removing the $2\omega_0$ terms)}
\end{align}
This is a rotation of the vector $x,y$ by $-\theta$, which is corrected with the transformation
\begin{align}
\begin{bmatrix}v_I\\v_Q\end{bmatrix} 
&=\begin{bmatrix}r_{11}&r_{12}\\r_{21}&r_{22}\end{bmatrix}
	\begin{bmatrix}v'_I\\v'_Q\end{bmatrix}\\[0em]
&= \begin{bmatrix}\cos\theta&-\sin\theta\\\sin\theta&\cos\theta\end{bmatrix}
	\begin{bmatrix}v'_I\\v'_Q\end{bmatrix}
\end{align}
Hence we have found Eq.~\req{eqn:cal-det-fix-theta-r}, QED.

\subsection{Accuracy considerations}
The above described process moves the measurement uncertainty from the RF/microwave section to the near-dc section, where the instrument uncertainty can be made negligible.  Additionally, we remark the following facts.
\vspace{-1ex}
\begin{itemize}\addtolength{\itemsep}{-1.2ex}
\item Modern lock-in amplifiers measure the real and imaginary part of the input signal after in-phase and quadrature down-conversion to dc.  Hence, the measurement of the small quadrature defect $\psi$ is actually a null measurement.  This is definitely true if the process is reiterated for best accuracy.

\item Using a digital lock-in, the lock-in internal quadrature reference relies on sampling, and ultimately on the internal clock.  This is free from the phase error of an analog detector.

\item The measurement of the small gain error $\epsilon$ is a differential measurement, which is limited by noise, rather than by the calibration of the lock-in.

\end{itemize}

\section{Correction of the I-Q modulator}\label{cal:sec-fix-iqmod}
\begin{figure}[t]
\centering\namedgraphics{0.8}{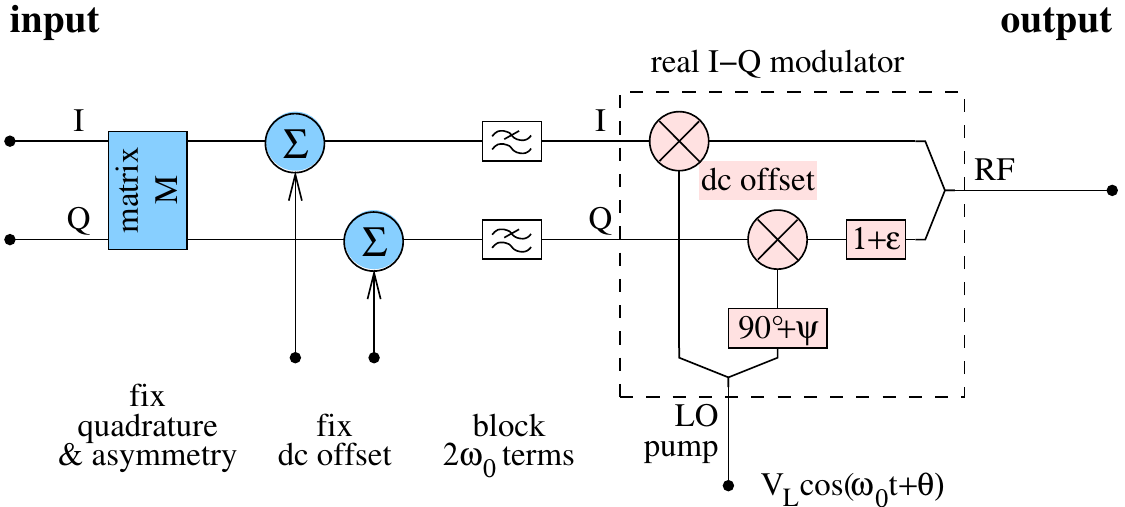}{\textwidth}
\caption{Correction of the real I-Q modulator.}
\label{fig:cal-modul-correct}
\end{figure}

This section describes how to compensate the defects of a I-Q modulator, turning it into a virtually ideal one.  The phase of the LO pump is still let arbitrary, as it results from the electrical layout.  The method, shown in Fig.~\ref{fig:cal-modul-correct}, is similar to that used to fix the I-Q detector.   

\subsection{Parameter measurement}
\begin{figure}[t]
\centering\namedgraphics{0.8}{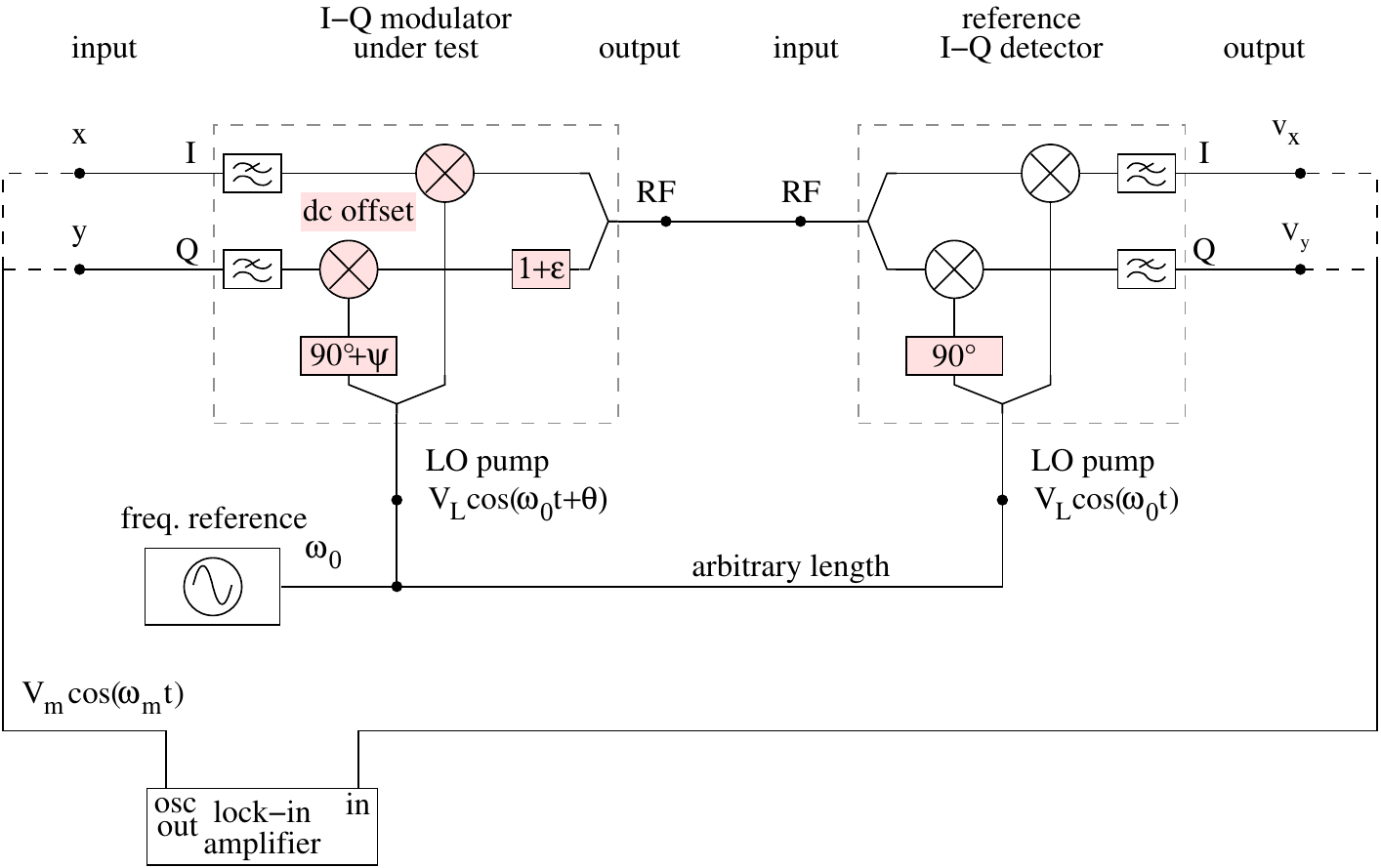}{\textwidth}
\caption{Measurementof the orthogonality and symmetry defect of a real I-Q modulator.}
\label{fig:cal-iq-channel}
\end{figure}

First, the LO crosstalk is eliminated by adding an appropriate dc offset, after inspection with a spectrum analyzer.  Afterwards, the matrix $M$ corrects for the orthogonality defect $\psi$ and for the gain asymmetry $\epsilon$.  The method to obtain the elements of $M$ from the measurements is shown in Figure~\ref{fig:cal-iq-channel}, and described underneath.

Feeding a signal 
\begin{align}
x &= V_m\cos(\omega_mt)\\
y &= 0
\intertext{at the modulator inputs, we calculate $\theta$ after measuring the output signals}
\label{eqn:cal-iq-mod-vxa}
v_x &= \frac12 V_m\cos(\omega_mt) \cos\theta\\
\label{eqn:cal-iq-mod-vya}
v_y &= \frac12 V_m\cos(\omega_mt) \sin\theta~.
\end{align}
Similarly, feeding a signal 
\begin{align}
x &= 0\\
y &= V_m\cos(\omega_mt)
\intertext{at the modulator inputs, we measure the output signals}
\label{eqn:cal-iq-mod-vxb}
v_x &= \frac12 V_m(1+\epsilon)\cos(\omega_mt)\sin(\theta+\psi)\\
\label{eqn:cal-iq-mod-vyb}
v_y &= \frac12 V_m(1+\epsilon)\cos(\omega_mt)\cos(\theta+\psi)~, 
\end{align}
from which we calculate $\epsilon$ and $\psi$.

It is to be remarked that the use of small signals of the form $V_m\cos(\omega_mt)$ allows the accurate measurement of $v_x$ and $v_y$ with a lock-in amplifier.  

A variant of this method consists of inserting a phase shifter in the RF line between the modulator and the detector, and to adjust the phase for $\theta=0$.  After eliminating $\theta$ with a null measurement, Eq.~\req{eqn:cal-iq-mod-vxb}-\req{eqn:cal-iq-mod-vyb} provide a more accurate estimation of $\epsilon$ and $\psi$.  The phase shifter should be introduced in the RF line, rather than in the LO pump of the modulator or of the detector, because these devices have a residual sensitivity to the pump power, and the phase shifter introduces a loss.

\subsection{Fixing the defects}
The system, as shown in Fig.~\ref{fig:cal-iq-channel} is equivalent to a channel described by the relationship
\begin{align}
\begin{bmatrix}v_x\\v_y\end{bmatrix} &=
\begin{bmatrix}c_{11}&c_{12}\\c_{21}&c_{22}\end{bmatrix}
\begin{bmatrix}x\\y\end{bmatrix}
\end{align}
where the matrix $C$ is 
\begin{align}
C &= \begin{bmatrix}\frac12 \cos\theta &-\frac12 \sin(\theta+\psi)\\[1ex]
	 \frac12 \sin\theta &\frac12 (1+\epsilon)\cos(\theta+\psi)\end{bmatrix}
\end{align}
Hence, the system is corrected by inserting the matrix $C^{-1}$ (the inverse of $C$) at the input 
\begin{align}
C^{-1} &= \begin{bmatrix}
\displaystyle\frac{2\cos(\theta+\psi)}{\cos\psi} 
	&\displaystyle\frac{2\sin(\theta+\psi)}{\cos\psi}\\[2ex]
\displaystyle-\frac{2\sin\theta}{(1+\epsilon)\cos\psi} 
	 &\displaystyle\frac{2\cos\theta}{(1+\epsilon)\cos\psi}
\end{bmatrix}
\\[1em]
&= \frac{2}{\cos\psi}
\begin{bmatrix}
\cos(\theta+\psi) &\sin(\theta+\psi)\\[1ex]
	 \displaystyle-\frac{\sin\theta}{1+\epsilon} 
	 &\displaystyle\frac{\cos\theta}{1+\epsilon}
\end{bmatrix}
\end{align}
so that $C^{-1}C=I$ (identity matrix).
As we have still not defined the measurement layout, there is no point in correcting $\theta$.  Thus we set $\theta=0$ in $C^{-1}$.  
The matrix $C^{-1}|_{\theta=0}$ is denoted with $M$
\begin{align}
M &= 2\begin{bmatrix}1&\displaystyle\frac{\sin\psi}{\cos\psi}\\[2ex]
	0 &\displaystyle\frac{1}{(1+\epsilon)\cos\psi}\end{bmatrix}
\end{align}
In summary, the matrix $M$ fixes $\epsilon$ and $\psi$, and lets the pump phase $\theta$ arbitrary.

\section{Building the reference AM-PM modulator}\label{cal:sec-build-ref-mod}
As a result of Section \ref{cal:sec-fix-iqmod} we have a virtually ideal I-Q modulator to be inserted in the reference AM-PM modulator (Fig.~\ref{fig:cal-modulator}).  Yet, the electrical layout introduces constant arbitrary phase shifts, which we represent as a single phase $\theta$ at the LO input of the I-Q modulator.  In this section we measure and correct $\theta$.

\subsection{Parameter measurement}
We first observe that phase modulators have a residual sensitivity to amplitude modulation and noise, while the power detector is insensitive to phase modulation and noise, provided the video bandwidth is large.  Therefore, we use a power detector to set $\theta$ for the I-Q modulator to add pure phase modulation to the carrier when it receives a signal at the $Q$ input.  This id detected as a null of amplitude modulation at the detector output.  Having corrected the geometry errors of the I-Q modulator, the channel $I$ adds pure amplitude modulation to the carrier.  

\begin{figure}[t]
\centering\namedgraphics{0.8}{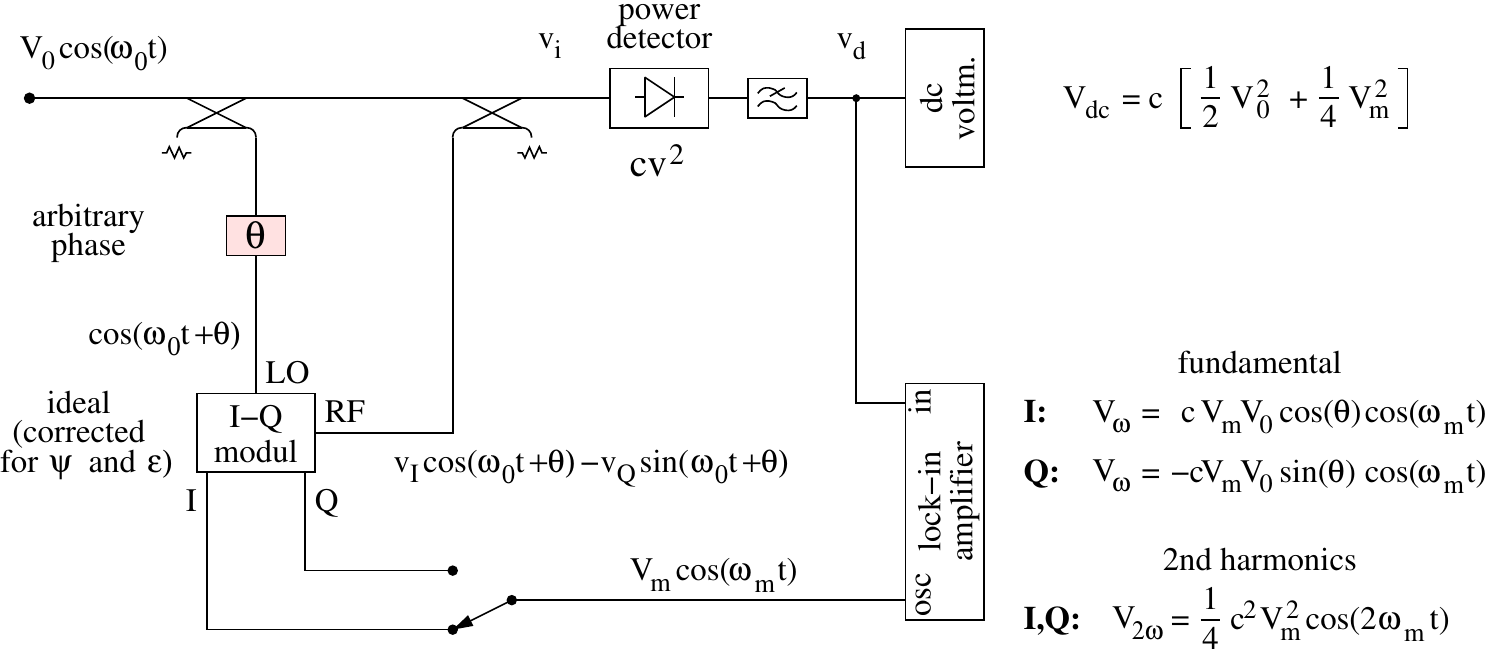}{\textwidth}
\caption{Setting the phase of the modulator pump.}
\label{fig:cal-meas-ampm-theta}
\end{figure}
Fig.~\ref{fig:cal-meas-ampm-theta} shows the method to measure $\theta$.
The I-Q pump is $\cos(\omega_0t+\theta)$, for the I-Q RF output signal is $v_I\cos(\omega_0t+\theta)-v_Q\sin(\omega_0t+\theta)$.  This is added to the carrier $V_0\cos(\omega_0t)$.  Setting the switch to $I$, we get
\begin{align}
v_I(t) = V_m\cos(\omega_mt) \qquad\text{and}\qquad v_Q(t) = 0
\end{align}
at the input of the I-Q modulator, and
\begin{align}
v_i(t) &= V_0\cos(\omega_0t) + V_m\cos(\omega_mt)\:V_0\cos(\omega_0t+\theta)
\end{align}
at the detector input.  The detected signal is
\begin{align}
\label{eqn:cal-measured-vd-x}
v_d(t) &= \frac12 V_0^2 + \frac14 V_m^2
	+ V_0V_m\cos(\theta)\cos(\omega_mt)
	+ \frac14 V_m^2 \cos(2\omega_mt)
\end{align}
Similarly, setting the switch to $Q$ we get
\begin{align}
v_I(t) = 0 \qquad\text{and}\qquad y(t) = v_Q\cos(\omega_mt) 
\end{align}
at the input of the I-Q modulator, and
\begin{align}
v_i(t) &= V_0\cos(\omega_0t) - V_m\cos(\omega_mt)\:V_0\sin(\omega_0t+\theta)
\end{align}
at the detector input.  The detected signal is
\begin{align}
\label{eqn:cal-measured-vd-y}
v_d(t) &= \frac12 V_0^2 + \frac14 V_m^2
	- V_0V_m\sin(\theta)\cos(\omega_mt)
	+ \frac14 V_m^2 \cos(2\omega_mt)
\end{align}
The lock-in amplifier selects the $\omega_m$ term in Equations~\req{eqn:cal-measured-vd-x} and \req{eqn:cal-measured-vd-y}, from which we calculate $\theta$.

\subsection{Fixing the defects}
The result of the measurement process is summarized as follows
\begin{align}
\begin{bmatrix}v_I\\v_Q\end{bmatrix}=\begin{bmatrix}V_m\\0\end{bmatrix}
&~~\Rightarrow~~~v_d=cV_0V_m\cos\theta
\intertext{and}
\begin{bmatrix}v_I\\v_Q\end{bmatrix}=\begin{bmatrix}0\\V_m\end{bmatrix}
&~~\Rightarrow~~~v_d=-cV_0V_m\sin\theta
\end{align}
which can be rewritten as the scalar product
\begin{align}
v_d	&= s^T\begin{bmatrix}v_I\\v_Q\end{bmatrix}\\
		&= \begin{bmatrix}\cos\theta & -\sin\theta\end{bmatrix}
			\begin{bmatrix}v_I\\v_Q\end{bmatrix} 
\end{align}
where the amplitude coefficients $cV_0V_m$ are dropped in order to simplify the formalism.

\begin{figure}[t]
\centering\namedgraphics{0.8}{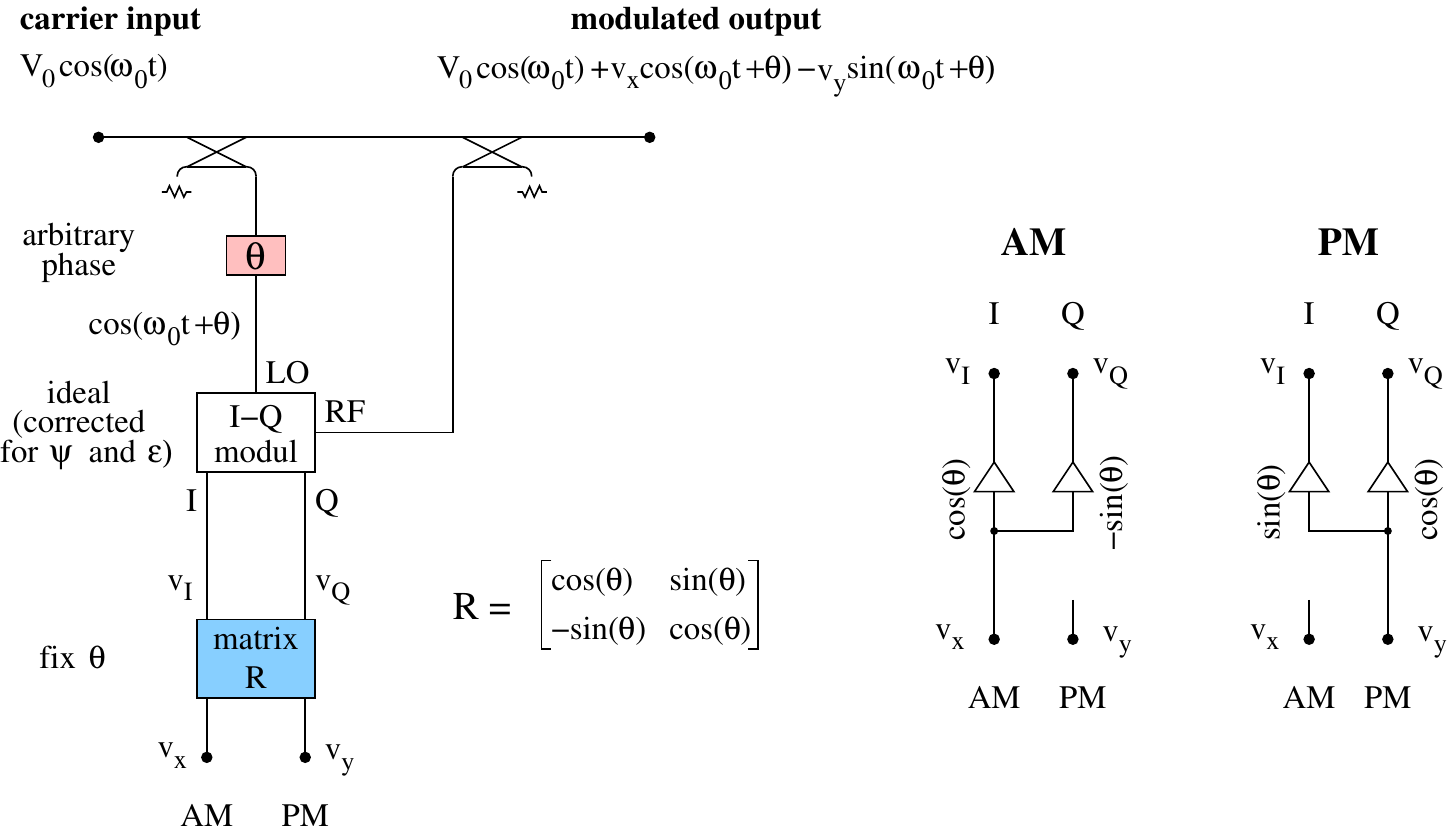}{\textwidth}
\caption{Setting the phase of the modulator pump.}
\label{fig:cal-build-ampm-mod}
\end{figure}
The arbitrary phase shift $\theta$ is compensated for by introducing a matrix $R$ at the input of the I-Q modulator (Fig.~\ref{fig:cal-build-ampm-mod})
\begin{align}
\begin{bmatrix}v_I\\v_Q\end{bmatrix} 
&=
\begin{bmatrix}r_{11} & r_{12}\\r_{11} & r_{12}\end{bmatrix}
\begin{bmatrix}v_x\\v_y\end{bmatrix} 
\end{align}
so that 
\begin{align}
v_d &=
\begin{bmatrix}\cos\theta & -\sin\theta\end{bmatrix}
\begin{bmatrix}r_{11} & r_{12}\\r_{11} & r_{12}\end{bmatrix}
\begin{bmatrix}v_x\\v_y\end{bmatrix} 
\end{align}
The matrix $R$ has to satisfy two conditions: when the input is $v_x=1$, $v_y=0$ the output is a modulation in phase with the carrier, which is detected
\begin{align}
\begin{bmatrix}\cos\theta & -\sin\theta\end{bmatrix}
\begin{bmatrix}r_{11} & r_{12}\\r_{11} & r_{12}\end{bmatrix}
\begin{bmatrix}1\\0\end{bmatrix} 
&=1\qquad\text{(AM)}
\intertext{and when the input is $v_x=0$, $v_y=1$ the output is a modulation in quadrature with the carrier, which is not seen by the power detector}
\begin{bmatrix}\cos\theta & -\sin\theta\end{bmatrix}
\begin{bmatrix}r_{11} & r_{12}\\r_{11} & r_{12}\end{bmatrix}
\begin{bmatrix}0\\1\end{bmatrix} 
&=0\qquad\text{(PM)}
\end{align}
The solution is
\begin{align}
R &=\begin{bmatrix}\cos\theta&\sin\theta\\-\sin\theta&\cos\theta\end{bmatrix}
\end{align}

\section{Calibration checklist}
The following list summarizes the sequence of actions for proper calibration.
\vspace{-1ex}
\begin{enumerate}\addtolength{\itemsep}{-1.2ex}
\item Correct the I-Q detector ($\epsilon$ and $\psi$) \ref{cal:sec-fix-iqdet}.
\item Correct the I-Q modulator ($\epsilon$ and $\psi$) \ref{cal:sec-fix-iqmod}.
\item Build the reference AM-PM modulator \ref{cal:sec-build-ref-mod}.
\vspace{-1ex}
\begin{itemize}\addtolength{\itemsep}{-0.7ex}
	\item chose the appropriate levels.
	\item fix $\theta$.
	\item measure the modulation depth.
\end{itemize}
\item Optional, needed for AM: fix $\theta$ of the detector.
\item Use the reference modulator to calibrate a measurement. system
\end{enumerate}

\section{Modulation depth and accuracy}
First we address a question left aside, why calibrating the system using a sinusoidal modulation instead of random or pseudorandom noise.  The reasons are detailed unerneath. 
\vspace{-1ex}
\begin{itemize}\addtolength{\itemsep}{-1.2ex}
\item Quadrature modulation is not true PM, it produces some AM\@.  Yet, the residual AM occurs at the frequency $2\omega_m$, which is rejected by a lock-in.
\item The noise bandwidth can be made narrow.
\item Can reject the background noise of the system. 
\end{itemize}
Though all these features can also be achieved with pseudorandom modulation, the sinusoidal modulation gives them in a simple and straightforward way.

\begin{figure}[t]
\centering\namedgraphics{0.8}{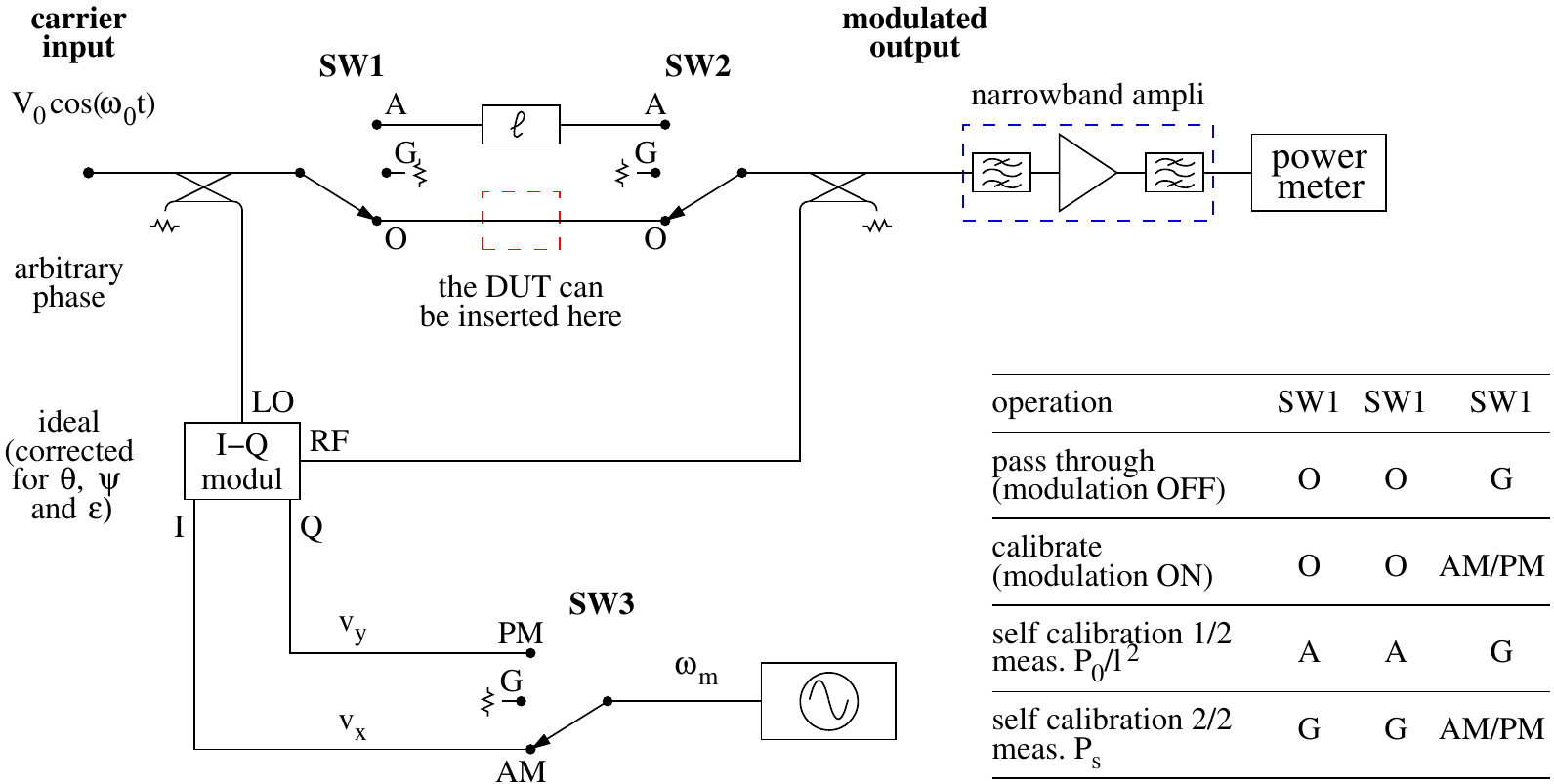}{\textwidth}
\caption{Measurement of the modulation depth.}
\label{fig:cal-modul-depth}
\end{figure}
Figure \ref{fig:cal-modul-depth} shows a variant of the reference modulator (first introduced in Fig.~\ref{fig:cal-modulator}), adapted to the measurement of the modulation depth.  The modulator is used in the following ways.
\begin{description}
\item[Pass through.] This is used to measure the noise of a DUT\@.  For highest accuracy of the signal path, the modulator is not removed.
This makes sense only if the modulator loss in series to the DUT can be tolerated. 
\item[Calibrate.]  Normal operation of the reference modulator, which provides a calibrated amplitude or phase modulation.
\item[Self calibration.] The reference modulator is calibrated in two steps, as detailed in Section~\ref{ssec:cal-mdepth}. 
\end{description}

\subsection{Phase and frequency modulation}
Sinusoidal phase and frequency modulations can be used as a reference of angle modulation.  Unfortunately, in (virtually) all practical cases this method does not provide simultaneously the reference unmodulated signal together with the modulated signal.

The basic phase modulation is 
\begin{align}
v(t) &= \cos\left[\omega_0t+\varphi(t)\right]
&&\text{(basic PM)}~.
\end{align}
In the presence of a sinusoidal modulation $\varphi(t)=m\sin(\omega_mt)$ of modulation index $m$, the signal writes
\begin{align}
v(t) &= \cos\left[\omega_0t+m\sin(\omega_mt)\right]
&&\text{(sinusoidal PM)}~,
\label{eqn:cal-sinus-pm}
\end{align}
which can be expanded as
\begin{align}
v(t) 
&= \cos(\omega_0t) \cos\left[m\sin(\omega_mt)\right]
	- \sin(\omega_0t) \sin\left[m\sin(\omega_mt)\right]
	\label{eqn:cal-sinus-pm-1}\\
&= J_0(m) \cos(\omega_0t) - 2J_1(m)\sin(\omega_mt) \sin(\omega_0t) + \ldots
\label{eqn:cal-sinus-pm-2}\\
&= J_0(m) \cos(\omega_0t)
	+ J_1(m)\sin(\omega_0+\omega_m)t - J_1(m)\sin(\omega_0-\omega_m)t
	\label{eqn:cal-sinus-pm-3}\\
&\simeq \cos(\omega_0t) 
	+ \tfrac12m\sin(\omega_0+\omega_m)t - \tfrac12m\sin(\omega_0-\omega_m)t
	\qquad\text{(small $m$)}~.
	\label{eqn:cal-sinus-pm-4}
\end{align}
Expanding the sinusoidal PM, we have used the following properties
\begin{align*}
&\cos\left[z\sin\theta\right]=J_0(z)\cos\theta+\ldots 
	\quad\text{and}\quad
	\sin\left[z\sin\theta\right]=2J_1(z)\sin\theta+\ldots~,\\
&\sin(a) \sin(b) = \tfrac12\left[\cos(a-b)-\cos(a+b)\right]~,\\
&J_0(z)=1+\ldots	\quad\text{and}\quad J_1(z)=\tfrac12z+\ldots
	\quad\text{for $m\rightarrow0$}~.
\end{align*}
Finally, equating the phase modulation $\varphi(t)=m\sin(\omega_mt)$ 
to the phase $\varphi(t)=\int(\Delta\omega)(t)\,dt$ obtained from the frequency modulation yields
\begin{align}
(\Delta\omega)(t)=(\Delta\omega)_p\cos(\omega_mt)
\qquad\text{(sinusoidal FM)}~,
\end{align}
and therefore
\begin{align}
m=\frac{(\Delta\omega)_p}{\omega_m}=\frac{(\Delta\nu)_p}{f_m}
\qquad\text{(modulation index)}~.
\end{align}

\subsubsection{Commercial synthesizers}
Using commercial synthesizers, the frequency modulation is preferred to the phase modulation because of easier calibration.  The DDS is a remarkable exception to this statement because the phase modulation can be obtained numerically by adding a term to the accumulator content, before addressing the look-up table.

The modulation index can be measured precisely measuring $(\Delta\nu)_p$ in static conditions with frequency counter, by switching a dc input and trusting the linearity of the VCO for small frequency deviation.

Another method, known as the \emph{Bessel null} method, consists of increasing $m$ until the carrier nulls.  From \req{eqn:cal-sinus-pm-2}, the carrier nulls at the first zero of  $J_0(m)$, which occurs at $m=2.405$.  Observing $v(t)$ with a spectrum analyzer, the carrier drops suddenly in a sharp interval.

A third method consists of the direct measurement of $m$ from the ratio of the carrier power $P_0$ divided by the sideband power $P_s$, using a spectrum analyzer.  
At small modulation index, \req{eqn:cal-sinus-pm-4} gives 
\begin{align}
\frac{P_s}{P_0}=\frac{1}{4}m^2~.
\end{align}

\subsection{Measurement of the modulation depth}\label{ssec:cal-mdepth}
We measure the carrier power and the modulation power separately.
A reference attenuation $\ell$ is introduced in the carrier path, so that the carrier power $P_0/\ell^2$ and the DSB power are of the same order of magnitude. This ensures precise power ratio measurements even with commercial power meters.  
The modulation is 
\begin{align}
\label{eqn:cal-mdepth-phirms-a}
\varphi_\text{rms} = \sqrt{\frac{P_y}{P_0}}
\qquad\text{or}\qquad
\alpha_\text{rms} &= \sqrt{\frac{P_x}{P_0}}
\end{align}
where $P_x$ applies to AM, $P_y$ to $PM$.  The modulation power ($P_x$ or $P_y$) accounts for both sidebands.
A detailed explanation of the method, restrained to the phase modulation, is given underneath.  The extension to the amplitude modulation is trivial.

The quadrature-modulated signal can be written as
\begin{align}
v(t) &= V_0\cos(\omega_0t) 
- V_s\sin[(\omega_0+\omega_b)t] 
- V_s\sin[(\omega_0-\omega_b)t]\\
&= V_0\cos(\omega_0t) 
+ 2V_s\sin(\omega_0t) \cos(\omega_bt)
\end{align}
The phase modulation $\varphi(t)$ associated to the above signal is  
\begin{align}
\varphi(t) 
&= \arctan\left[2\frac{V_s}{V_0}\cos(\omega_bt)\right]\\ 
&\simeq 2\frac{V_s}{V_0}\; \cos(\omega_bt) 
\qquad\text{linearized for small $V_s/V_0$}
\intertext{thus}
\varphi_\text{rms}&=\sqrt2\,\frac{V_s}{V_0}
\label{eqn:cal-mdepth-phirms-b}
\end{align}
The carrier power is $P_0=\smash{\frac{V_0^2}{2R_0}}$, where $R_0$ is the load resistance.  Similarly, the power of each sideband is $P_s=\smash{\frac{V_s^2}{2R_0}}$, thus $P_y=\smash{2\frac{V_s^2}{2R_0}}$ for both sidebands.  Hence, the power ratio $\frac{P_m}{P_0}$ is
\begin{align}
\frac{P_y}{P_0} 
= 2\frac{V_s^2}{2R_0} \times \frac{2R_0}{V_0^2} 
= 2\frac{V_s^2}{V_0^2}~. 
\end{align}
By virtue of Eq.~\req{eqn:cal-mdepth-phirms-b}, the above is equal to $\left<\varphi^2\right>$, which proves Equation~\req{eqn:cal-mdepth-phirms-a}.

\subsection{Mixer response and linearity}\label{ssec:cal-mixer-linearity}
The sole mixer of our interest in this article is the double-balanced mixer based on a ring of Schottky diodes (see \cite{rubiola06arxiv-tutorial-on-mixers} for a tutorial).  The mixer used as an analog modulator is expected to respond linearly
\begin{align}
v_{RF} &= a\,i_{IF}
\end{align}
where $a$ is an appropriate coefficient.  Actual mixers deviate from linearity, and saturate at an output power of some 1 mW (0 dBm).  We measured a mixer Mini Circuits ZFM-2 driven at the IF input by a current generator in the following conditions:\par
\begin{center}\begin{tabular}{ll}
LO frequency	& 100 MHz\\
LO power		& 5 mW (7 dBm)\\
RF load		& 50 $\Omega$\\
IF resistance	& 1 k$\Omega$
\end{tabular}\end{center}\par
\begin{figure}[t]
\centering\namedgraphics{0.8}{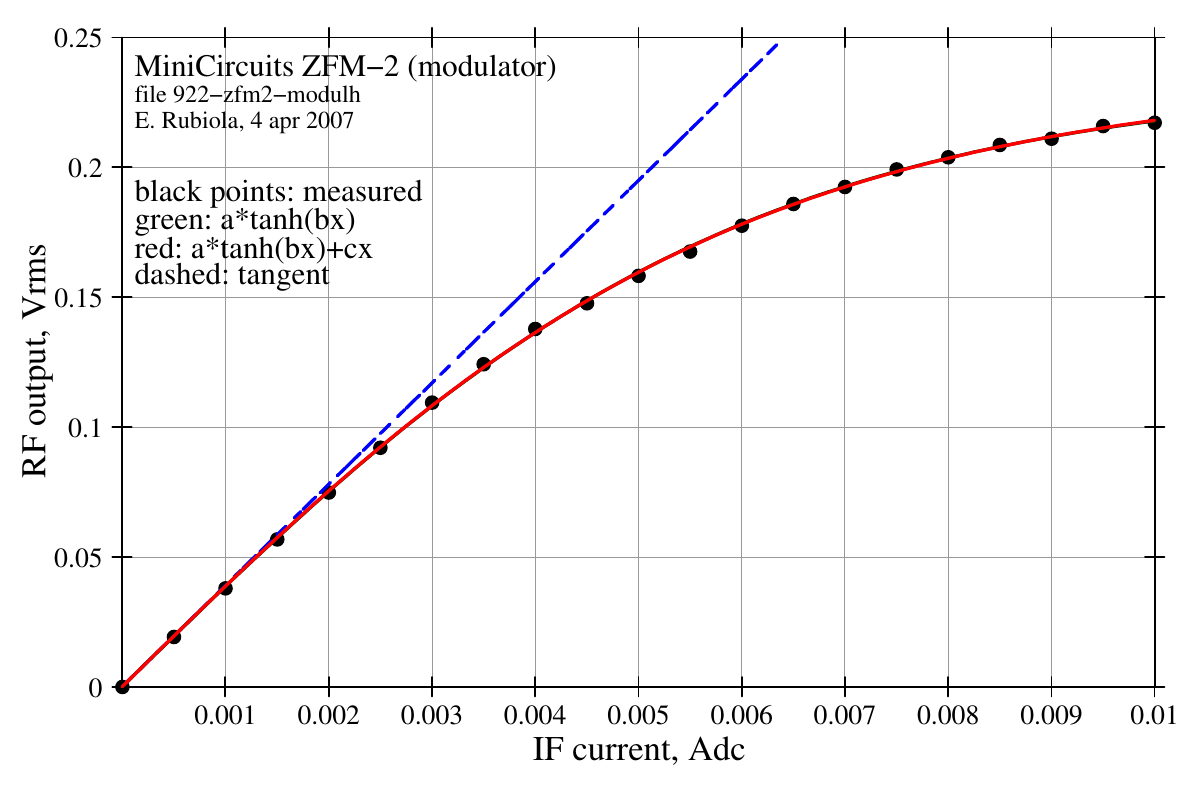}{\textwidth}\\[1em]
\centering\namedgraphics{0.8}{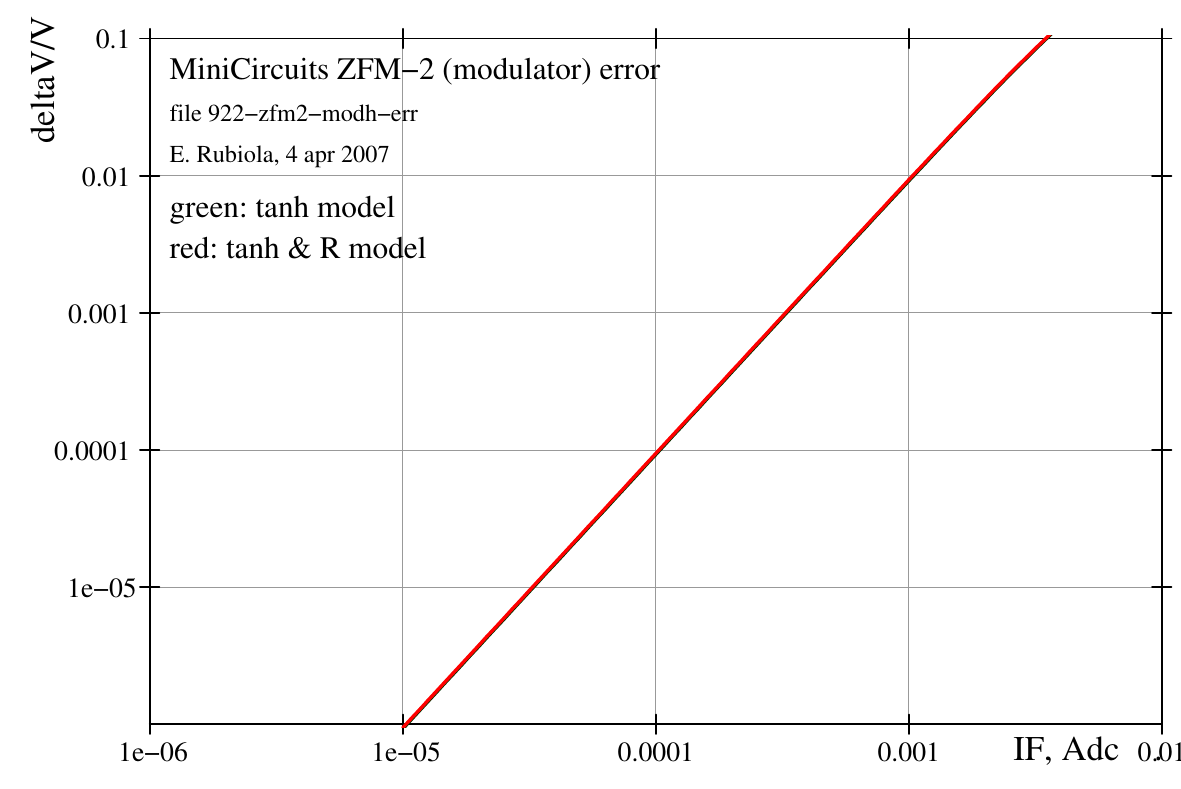}{\textwidth}
\caption{Double-balanced mixer used as an analog modulator.  Top: experimental data, polynomial fit, and small-signal (linear) approximation.  Bottom: discrepancy between model and linear approximation.}
\label{fig:922-zfm2-modulh}
\end{figure}
\begin{figure}[t]
\centering\namedgraphics{0.8}{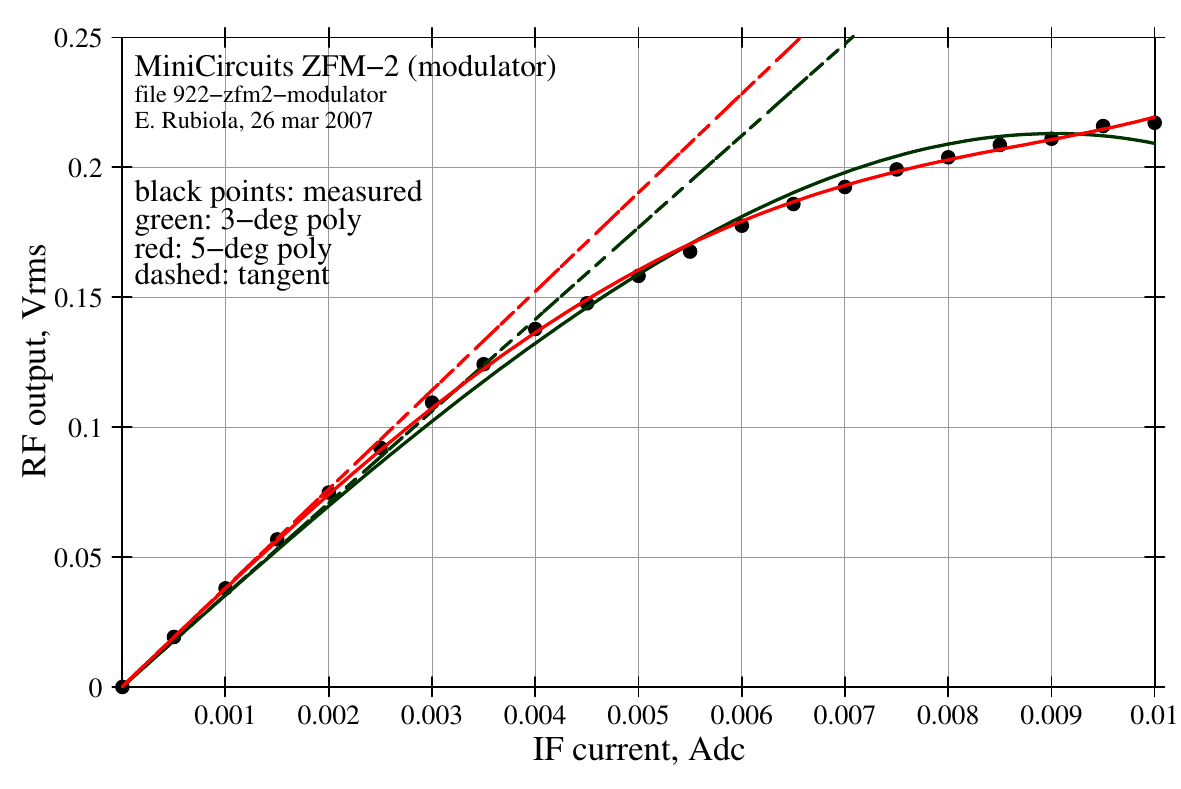}{\textwidth}\\[1em]
\centering\namedgraphics{0.8}{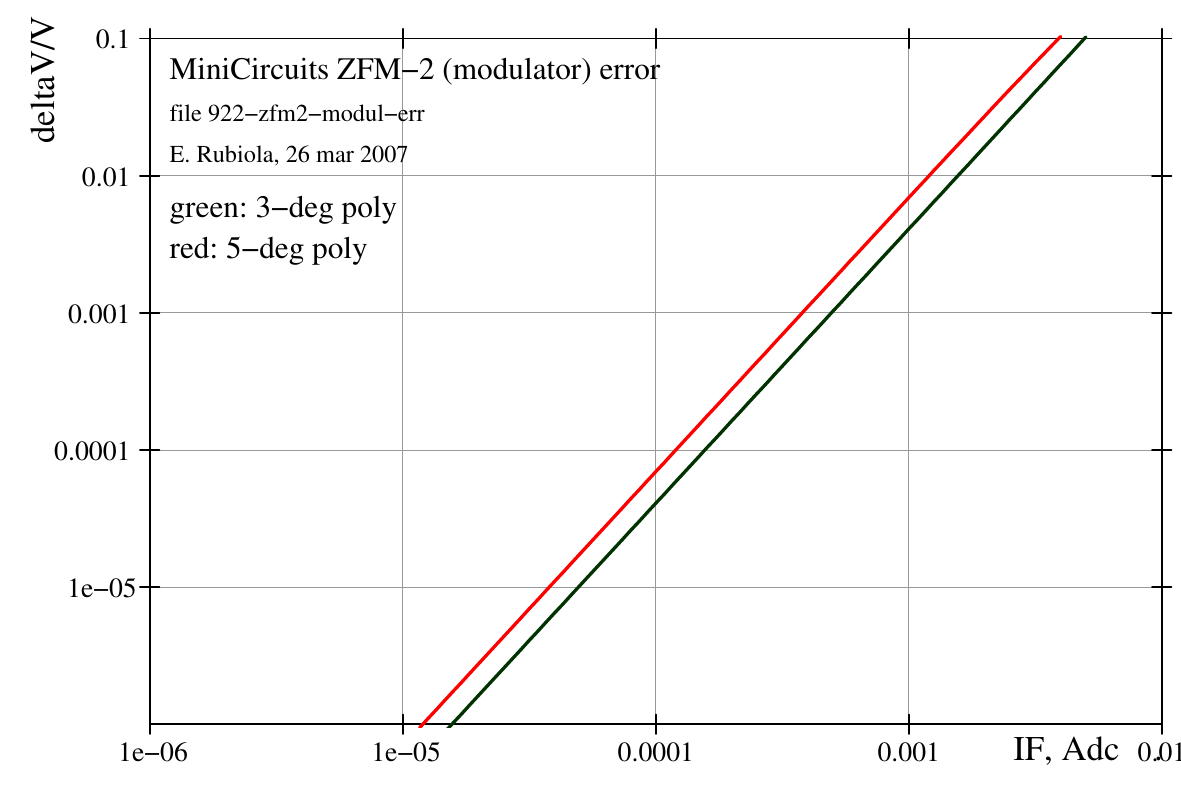}{\textwidth}
\caption{Double-balanced mixer used as an analog modulator.  Top: experimental data, 3rd and 5th degree polynomial fit, and small-signal (linear) approximation.  Bottom: discrepancy between model and linear approximation.}
\label{fig:922-zfm2-modulator}
\end{figure}

 
Due to the exponential behavior of the diodes and to the even symmetry, similar to that of a differential pair, it is natural to model the mixer as  
\begin{align}
v_{RF} &= a_1 \tanh(a_2i_{IF})	&&\text{pure $\tanh(x)$ model}
\label{eqn:cal-tanh}
\intertext{or as}
v_{RF} &= a_1 \tanh(a_2i_{IF})+a_3i_{IF}	&&\text{$\tanh(x)$ model with dissipation}
\label{eqn:cal-tanh-r}
\end{align}
The second model differs from the first by the insertion of a series resistance, which accounts for the diode and transformer resistance.  The experimental data and the least-square approximations are shown in Figure~\ref{fig:922-zfm2-modulh}.
For the sake of completeness, we also try a polynomial fit (Fig.~\ref{fig:922-zfm2-modulator}).  Table~\ref{tab:cal-fit-coefficients} summarizes the coefficients of the least-square fits.

Looking at the above models, we notice that the $\tanh(x)$ model is superior to the polynomial fit, and that adding the dissipative term has a small beneficial effect.     
Table~\ref{tab:cal-mixer-linearity}
gives an idea of the error $\Delta v_{RF}/v_{RF}$ if the the full Equation \req{eqn:cal-tanh} or \req{eqn:cal-tanh-r} is replaced with the tangent.

\begin{table}[t]
\caption{Linearity defect [Eq.~\req{eqn:cal-tanh-r}] of a double-balanced mixer used as a modulator.}
\label{tab:cal-mixer-linearity}
\begin{center}\begin{tabular}{|cc|ccc|}\hline
\boldmath$i_{IF}$	&\boldmath$\Delta v_{RF}/v_{RF}$& $v_{RF}$& \multicolumn{2}{c|}{$P_{RF}$} \\\hline
0.1	& $10^{-4}$	& 3.91	&0.305	& $-35.2$\\
0.316& $10^{-3}$	& 12.4	&3.05	& $-25.2$\\
1	& $10^{-2}$	& 39.1	& 30.5 	& $-15.2$\\\hline
mA	&(dimensionless) & mV$_\text{rms}$	&$\mu$W&dBm\\\hline
\end{tabular}\end{center}
\end{table}

We expect that the model \req{eqn:cal-tanh} or \req{eqn:cal-tanh-r} is a factor 10 more accurate than the linear approximation, that is, $10^{-3}$ at $i_{IF}=1$ mA and $10^{-5}$ at $i_{IF}=100$ $\mu$A\@.
We expect uncertainty and errors of the same order when the mixer is used as a synchronous detector.
Of course, this is the accuracy of the model, which does \emph{not} include the uncertainty of the instruments used to measure the mixer.

\begin{table}[t]
\caption{Coefficients of the numerical fits for a ZFM-2 mixer used as a modulator (see Fig.~\ref{fig:922-zfm2-modulh} and \ref{fig:922-zfm2-modulator}).}
\label{tab:cal-fit-coefficients}
\centering\begin{minipage}{0.85\textwidth}
\vspace{1ex}\begin{footnotesize}\begin{verbatim}
--- polynomial approx, RMS voltage -------------------------------
a1       a3        a5       a7         a9        slope    residual
3.536e1  -1.444e5                                3.536e1  2.091e-2
3.805e1  -2.641e5  1.030e9                       3.805e1  6.516e-3
3.902e1  -3.485e5  2.814e9  -1.060e13            3.902e1  3.835e-3
3.927e1  -3.833e5  4.129e9  -2.876e13  8.281e16  3.927e1  3.692e-3
--- polynomial approx, PEAK voltage ------------------------------
a1       a3        a5       a7         a9        slope    residual
5.001e1  -2.042e5                                5.001e1  2.957e-2
5.381e1  -3.735e5  1.456e9                       5.381e1  9.216e-3
5.518e1  -4.928e5  3.979e9  -1.499e13            5.518e1, 5.423e-3
5.553e1  -5.421e5  5.839e9  -4.067e13  1.171e17  5.553e1  5.221e-3
--- hyperbolic-tangent approx, RMS voltage -----------------------
a1       a2        a3                            slope    residual
2.340e-1 1.666e2                                 3.898e1  4.043e-3
2.271e-1 1.695e2   5.834e-1                      3.907e1  4.013e-3
--- hyperbolic-tangent approx, PEAK voltage ----------------------
a1       a2        a3                            slope    residual
3.310e-1  2.356e2                                5.513e1  5.718e-3
3.212e-1  2.397e2  8.250e-1                      5.526e1  5.676e-3
------------------------------------------------------------------
\end{verbatim}\end{footnotesize}
\end{minipage}
\end{table}

\subsection{Power levels and signal-to-noise ratio (SNR)}\label{ssec:cal-snr}
\begin{figure}[t]
\centering\namedgraphics{0.8}{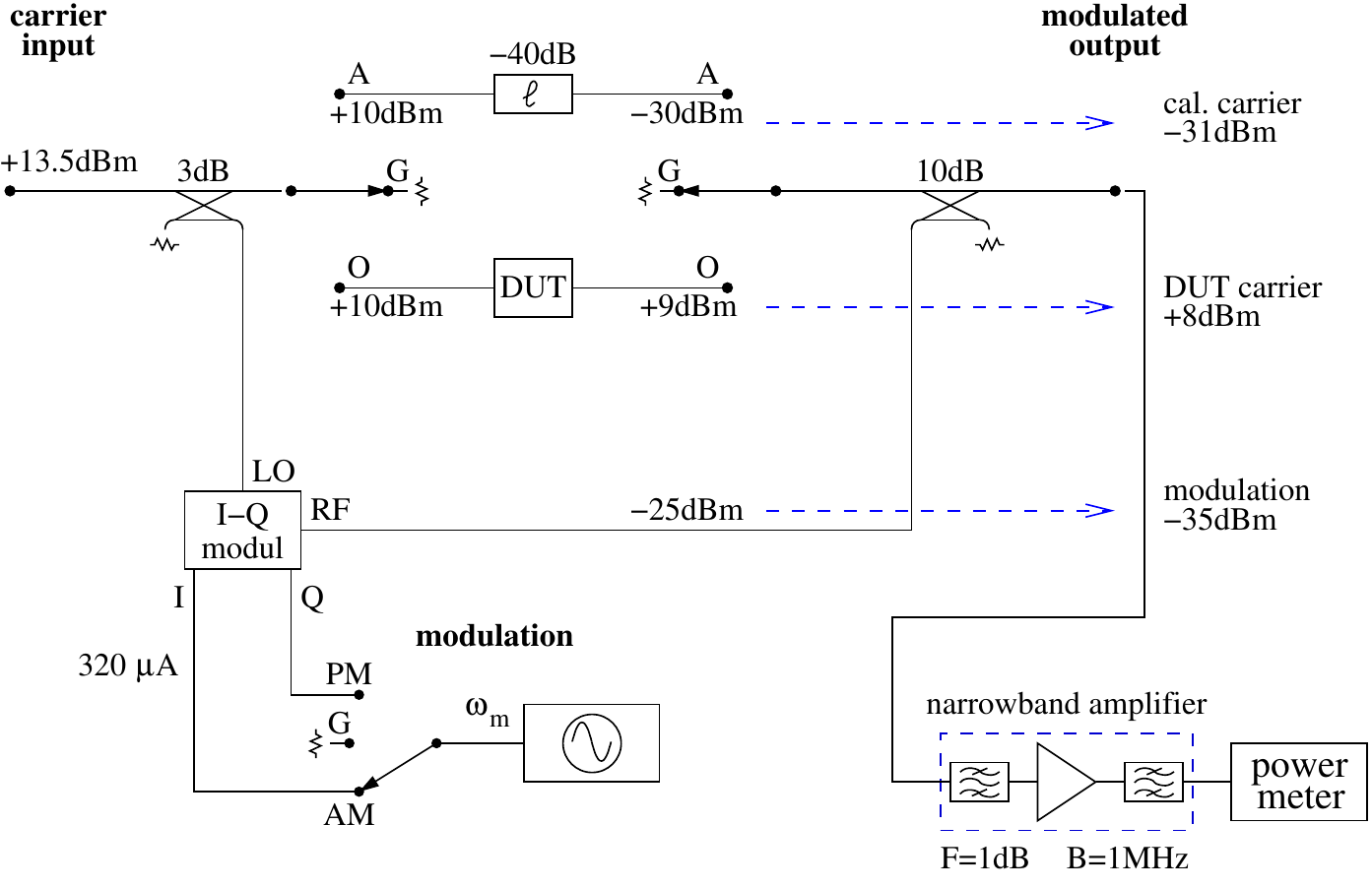}{\textwidth}
\caption{Tentative estimation of signal levels and SNR.}
\label{fig:cal-snr}
\end{figure}
The choice of power levels depends on the application, and chiefly on the power of the signal to be calibrated.  Figure~\ref{fig:cal-snr} (derived from Fig.~\ref{fig:cal-modul-depth}) provides a first example, in which the DUT is suitable to a power of 10 mW ($+10$ dBm) and shows an attenuation of 1 dB\@.  The directional coupler in which the modulation is injected shows an attenuation of 1 dB (intrinsic and dissipative).  Thus $P_0=6.3$ mW ($+8$ dBm).
We measure $P_0$ by inserting a reference 40 dB attenuator (round value), so that $P_0/\ell^2=800$ nW ($-31$ dBm).  In this 
We try $i_{IF}=316$ $\mu$A as a compromise between linearity and SNR\@.  In this conditions the modulation power is $P_y=316$ nW, after 10 dB attenuation of the directional coupler.  
The equivalent noise power at the amplifier input is $P_N=FkTB$.  Assuming a noise figure $F=1.25$ (1 dB) and a noise bandwidth $B=1$ MHz, we get $P_N=5{\times}10^{-15}$ W, thus a signal-to-noise ratio
\begin{align}
\text{\emph{SNR}}=\frac{P_y}{P_N}
=\frac{3.05{\times}10^{-7}}{5.04{\times}10^{-15}}
=6.05{\times}10^8\qquad\text{(87.8 dB)}
\end{align}
It is to be remarked that $B$ impacts only on the SNR, thus on the assessment of the uncertainty, \emph{not} on the modulation depth.  A narrower $B$ can be used, provided the bandpass region is flat enough to attenuate the carrier and the sidebands equally. 
Table~\ref{tab:cal-mixer-snr} compares some calibration scenarios.

\begin{table}[t]
\caption{Expected SNR of the scheme shown in Fig.~\ref{fig:cal-snr}.}
\label{tab:cal-mixer-snr}
\begin{center}\begin{tabular}{|c|cc|cc|}\hline
\boldmath$i_{IF}$ &\multicolumn{2}{c|}{$P_{RF}$} &
\multicolumn{2}{c|}{\bfseries SNR}\\\hline
0.1	& 0.305	& $-35.2$	&$7.79{\times}10^3$&77.8\\
0.316& 3.05	& $-25.2$	&$2.46{\times}10^4$&87.8\\
1	& 30.5 	& $-15.2$ &$7.79{\times}10^4$&97.8\\\hline
mA	&$\mu$W&dBm & (dimensionless) & dB\\\hline
\end{tabular}\end{center}
\end{table}

\subsection{Tentative estimation of the uncertainty}
Though the method relies only on null and ratio measurements, having primary-metrology facility for power and impedance on site helps in achieving the highest accuracy.  
We attempt to estimate the uncertainty under the hypothesis that all the calibration is based on commercial instruments and that no primary metrology is available on site, which is our case.
Table~\ref{tab:cal-uncertainty} summarizes the uncertainties, keeping on conservative values.

\begin{table}[t]
\caption{Tentative estimation of the uncertainty, based on commercially available component and instruments.  The values given below are thought to be conservative.}
\begin{center}\begin{tabular}{lcc}\hline\hline
parameter and conditions	&\multicolumn{1}{c}{value}&\\\hline\hline
power ratio measurement	& $11.6{\times}10^{-3}$	& (0.1 dB)\\
(commercial power meter)	&					& \\\hline
RF path				&  $23{\times}10^{-3}$	& (0.2 dB)\\
(couplers, cables etc.)	& &\\\hline
reference attenuator& $5.8{\times}10^{-3}$	& (0.05 dB)\\
(Weinschel, 40 dB)		& &\\\hline
mixer and detector linearity& $1.0{\times}10^{-3}$	& \\
Section \ref{ssec:cal-mixer-linearity}&&\\\hline
null measurements		& $1.0{\times}10^{-3}$	& \\
(commercial lock-in, 10 bit)& 					&\\\hline
signal-to-noise ratio		& $1.0{\times}10^{-3}$	& \\
Section \ref{ssec:cal-snr}	&&\\
\hline\hline
\multicolumn{1}{r}{worst case total}	& $43.6{\times}10^{-3}$	&(0.37 dB)\\
\multicolumn{1}{r}{rms total}	& $26.5{\times}10^{-3}$	&(0.23 dB)\\
\hline\hline
\end{tabular}\end{center}
\label{tab:cal-uncertainty}
\end{table}

\section{Application to the bridge (interferometric) instruments}
\begin{figure}[t]
\centering\namedgraphics{0.9}{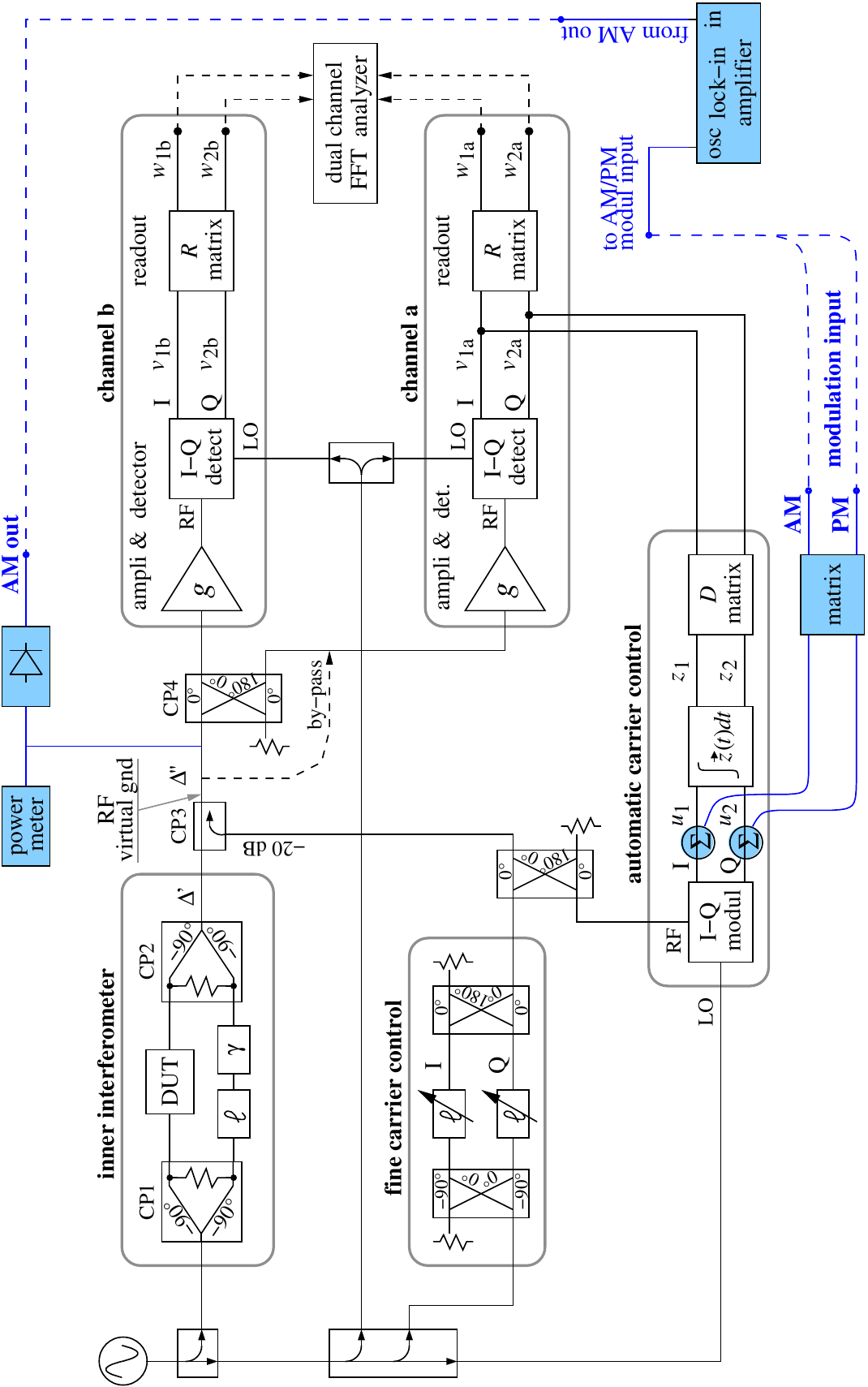}{\textwidth}
\caption{The calibration method fits well into the bridge (interferometric) noise measurement instruments.  The notation is that of the reference \cite{rubiola2002rsi-matrix}, where the dual bridge is published.  In color, the changes needed for calibration.}
\label{fig:cal-dual-bridge}
\end{figure}
At the end of the conceptual work I realized that my bridge (interferometric) instruments for the measurement of AM and PM noise \cite{rubiola2002rsi-matrix} contain almost all the blocks needed for the calibration.  Therefore, the calibration process can be added with a minimum of changes.  This concept is shown in Figure \ref{fig:cal-dual-bridge}, where the new blocks (in color) are added to the instrument.

\section{Where these ideas come from}
Though at superficial sight our scheme has some similarity to a scheme patented by F. L. Walls at NIST \cite{walls1992patent-calibration}, theory of operation, calibration procedure, and use are radically different.
The key idea of synthesizing a modulated signal by adding carrier and sidebands is so old that I can't track back the original idea.

I wish to conclude this Section with a warning against patents.   The SI units, by definition, can not be patented.  
In the official documents of \emph{La convention du m\`etre} and its child BIPM it is written clearly that the SI units are \emph{defined} and \emph{designed for} them to be \emph{free} and \emph{accessible}.  Some readers could remember that one of the major objection against the definition of the second based on the Earth revolution around the Sun, in vigor in 1964--68, was that it was not sufficiently accessible.

\section*{Acknowledgements}
\addcontentsline{toc}{section}{Acknowledgements}
I am indebted with John Dick (JPL, now retired), Michele Elia (Politecnico di Torino) and Vincent Giordano (FEMTO-ST) for numerous discussions and a wealth of suggestions.  Vincent Giordano also supported me for more than ten years.
I wish to thank Patrice Salzenstein (project leader, FEMTO-ST) for managing and for supporting my ideas; Yannick Gruson and Simon Pascaud (FEMTO-ST) for technical support.  

The main support for this project comes from the Laboratoire National de M\'{e}trologie et d'Essai under contract no.~LNE/DRST\,08\,7\,002.  I thank St\'{e}phane Gille for patient and efficient management on the LNE side. 

Partial support and help also comes from the Elisa Project (cryogenic sapphire oscillator), ESA contract no.~20135/06/D/MRP\@.

\def\bibfile#1{/Users/rubiola/Documents/articles/bibliography/#1}
\addcontentsline{toc}{section}{References}
\bibliographystyle{abbrv}
\bibliography{\bibfile{ref-short},%
              \bibfile{references},%
              \bibfile{rubiola}}

\end{document}